\documentclass[a4paper]{elsarticle}

\usepackage[english]{babel}
\usepackage[utf8x]{inputenc}
\usepackage[T1]{fontenc}

\usepackage[a4paper,top=3cm,bottom=2cm,left=3cm,right=3cm,marginparwidth=1.75cm]{geometry}

\usepackage{amsmath}
\usepackage{amsthm}
\usepackage{amssymb}
\usepackage{mathtools}
\usepackage{bm}
\usepackage{graphicx}
\usepackage[colorinlistoftodos]{todonotes}
\usepackage[colorlinks=true, allcolors=blue]{hyperref}
\usepackage{epstopdf}
\usepackage{upgreek}

\usepackage{algpseudocode}

\usepackage{tabularx}
\newcolumntype{Y}{>{\centering\arraybackslash}X}

\usepackage{overpic}
\usepackage{sidecap}

\usepackage[framemethod=TikZ]{mdframed}

\usepackage{floatrow}
\usepackage{tikz}
\usepackage[font=footnotesize,labelfont=bf]{caption}

\usepackage[outline]{contour}
\contourlength{1.3pt}

\usepackage[export]{adjustbox}

\usepackage{subcaption}

\usepackage{algorithm}

\makeatletter
\def\ps@pprintTitle{%
	\let\@oddhead\@empty
	\let\@evenhead\@empty
	\def\@oddfoot{}%
	\let\@evenfoot\@oddfoot}
\makeatother

\newtheorem{theorem}{Theorem}[section]
\newtheorem{lemma}[theorem]{Lemma}

\theoremstyle{definition}

\newtheorem{example}{Example}[section]



\makeatletter
\newenvironment{procedure}[1][htb]{%
    \renewcommand{\ALG@name}{Procedure}
   \begin{algorithm}[#1]%
  }{\end{algorithm}}
\makeatother

\floatname{algorithm}{Optimization problem}

\makeatletter
\newcommand{\mathleft}{\@fleqntrue\@mathmargin1em}
\newcommand{\mathcenter}{\@fleqnfalse}
\makeatother

\def\aw{{\mathbf a}}
\def\bw{{\mathbf b}}
\def\cw{{\mathbf c}}
\def\dw{{\mathbf d}}
\def\ew{{\mathbf e}}
\def\gw{{\mathbf g}}
\def\nw{{\mathbf n}}

\def\Pw{{\mathbf P}}
\def\rw{{\mathbf r}}
\def\sw{{\mathbf s}}
\def\Sw{{\mathbf S}}

\def\vw{{\mathbf v}}
\def\xw{{\mathbf x}}

\begin{document}
\begin{frontmatter}


\title{Optimizing B-spline surfaces for developability and paneling architectural freeform surfaces}




\author[tuwien,evolute]{Konstantinos Gavriil\corref{cor}}
\ead{gavriil@evolute.at}
\author[evolute]{Alexander Schiftner}
\ead{schiftner@evolute.at}
\author[tuwien,kaust]{Helmut Pottmann}
\ead{pottmann@geometrie.tuwien.ac.at}

\address[tuwien]{Applied Geometry, Institute of Discrete Mathematics and Geometry,\\
Vienna University of Technology, Wiedner Hauptstrasse 8-10/104, A-1040 Vienna, Austria}
\address[evolute]{Evolute GmbH, Schwindgasse 4/10, 1040 Vienna, Austria}
\address[kaust]{Visual Computing Center, KAUST, Thuwal 23955-6900, Saudi Arabia}

\cortext[cor]{Corresponding author}

\begin{abstract}
Motivated by applications in architecture and design, we present a novel method for increasing the developability of a B-spline surface. We use the property that the Gauss image of a developable surface is 1-dimensional and can be locally well approximated by circles. This is cast into an algorithm for thinning the Gauss image by increasing the planarity of the Gauss images of appropriate neighborhoods. A variation of the main method allows us to tackle the problem of paneling a freeform architectural surface with developable panels, in particular enforcing rotational cylindrical, rotational conical and planar panels, which are the main preferred types of developable panels in architecture due to the reduced cost of manufacturing.
\end{abstract}


\end{frontmatter}


\section{Introduction}

Developable surfaces can be locally mapped to a planar domain without distortion. Since they can be constructed from an initial planar state without stretching or tearing, only by bending, they represent the shapes obtainable with thin materials like sheet metal or paper which do not stretch. These surfaces are of great interest to many applications. Areas like architecture, manufacturing and design take advantage of the cost-reduced manufacturing process that developables have.

Developable surfaces have been well studied in classical differential geometry. Developable, twice
differentiable surfaces are \emph{single curved}, meaning one of the principal curvatures is zero. Thus,
the Gauss curvature vanishes at every point. They are composed of special \emph{ruled} surfaces with a constant tangent plane at all points of a ruling. As the surface normal vectors along a ruling
agree, the Gauss image of a developable surface is 1-dimensional, i.e. a curve. 

 We base the main method in our paper on this property of the Gauss image. 
However, our focus is not on exact developability, but rather on 
{\it nearly developable} surfaces which we characterize by nearly curve-like Gauss images.
The motivation for our research is the fact that most materials allow for a little bit of stretching
and therefore developability needs not be satisfied to a high degree in a variety of applications.
In particular, we are interested in applications in architecture where various kinds of tolerances
can be exploited to reduce the production cost of freeform skins. Our work fits into a larger
research program on novel digital tools which consider key
aspects of function and fabrication, including material behavior, already in the early design
and digital modeling phase.  

\bigskip
\noindent\textbf{Previous work.} 
There is a vast amount of literature on developable surfaces, on their theory, their computational
design using various types of representations and on their appearance in numerous applications. 
We limit this discussion to three main areas which are most closely related to our work: (i) developable
Bezier and B-spline surfaces, (ii) discrete representations and nearly developable surfaces and (iii) their importance in paneling architectural surfaces. 

\smallskip\noindent {\em Developable Bezier and B-spline surfaces.} 
Lang and R\"oschel \cite{langroeschel92} expressed 
developability of rational, in particular polynomial B\'ezier surfaces
in a system of cubic equations. In general, this system cannot be solved
in a simple way, but in various special cases, explicit solutions 
have been derived (\cite{aumann:1991,aumann:2003,chusequin:2002,chuchen2004}).
One can avoid these nonlinear constraints by using the projectively dual representation,
where a developable is represented as the
envelope of its tangent planes. For details, we refer to  \cite[Section 6.2]{pottwall:2001}, 
but note that the dual representation is not sufficiently intuitive to be
suitable for interactive design. Moreover, it is difficult to control singularities. 
A combination of the primal and the dual representation has
been successfully employed for interactive design of developable NURBS
surfaces by Tang et al. \cite{chengcheng-dev-2015}. 

\smallskip\noindent {\em Discrete representations and nearly developable surfaces.} 
There are numerous papers which  model developable surfaces 
with triangle meshes; we just refer to a few of 
them \cite{frey:2003, mitanisuzuki:2004, roseetal:2007, tang:2004}. 
Jung et al. \cite{Jung:2015:SFD:2843519.2749458} improve on Decaudin at al.'s \cite{doi:10.1111/j.1467-8659.2006.00982.x} method that locally approximates neighborhoods around each mesh triangle with a cone.
Liu  et al. \cite{liu-2006-cm} treat developable surfaces as a limit case 
of meshes from planar quads. 
Solomon et al. \cite{solomon:2012} use a mesh approach to 
flexibly model the shapes achievable by bending and folding
a given planar domain without stretching or tearing. 
An elegant discrete model of developable surfaces is provided by special
quad meshes which discretize orthogonal nets of geodesics \cite{Rabinovich:DogNets:2018,Rabinovich:ShapeSpace:2018}. 

Nearly developable surfaces appear in connection with specific applications, e.g. modeling 
ship hulls \cite{perez:2007} and clothing \cite{chen2010} or segmenting 
meshes in geometry processing \cite{juliusetal:2005,yamauchietal:2005}.
Narain et al.\ \cite{Narain:2013:FCA} go beyond developability and present a technique for simulating plastic deformation in sheets of thin materials, such as crumpled paper, dented metal, and wrinkled cloth. 
Closely related to our work is a paper by Wang et al.\cite{Wang2004OnIT} on increasing developability of a trimmed NURBS surface, but our approach and applications differ significantly. 

Another very recent work with a strong connection to our research is the developable surface
flow by Stein et al. \cite{Steinetal:2018}. This flow is a gradient flow on the energy 
$\int_M \kappa_1^2\ dA$, $\kappa_1$ being the smallest principal curvature. It constructs piecewise
developable rather than globally developable surfaces as minimizers. The discrete model is based on triangulations whose
vertex stars dominantly lie in pairs of planes. One could say that the surface is locally
approximated by a pair of planes, their intersection representing the ruling direction.
In a similar spirit, our local approximations are of higher order, as discussed below. 
 Note that Stein et al. generate {\it piecewise} developable
surfaces, where the arising pattern of developable patches is a result of the geometric flow and
depends on the initial triangulation. We can increase developability of a single smooth surface without
the introduction of tangent discontinuities. We can also allow for  piecewise developable surfaces through an 
appropriate selection of knots and their multiplicities in the underlying B-spline surface, but our 
arrangements of developable patches are more restricted (and at the same time more
controlled) than the ones by Stein et al.

\smallskip\noindent {\em Paneling architectural surfaces.} Architectural surfaces need to be decomposed
into panels, which is a key process and largely responsible for a cost effective solution.
For an overview of the problems in this field we refer to \cite{pottmann-2015-ag}. In particular,
we point to the paneling solution of Eigensatz et al. \cite{pottmann-2010-pan}. It exploits
various tolerances at seams and a cost model for the production of panels of different geometric
types to suggest solutions within an optimization framework. The user provides the design surface and a suggested network of panel boundary curves, while the algorithm slightly adapts the design surface and network and optimally fills it with panels (patches). Our work can be considered as an extension
in the sense that the panel boundaries are also subject to optimization with the overall goal
of increasing developability of the individual panels. For developable and nearly developable surfaces in architecture, we further point to \cite{pottmann-2008-strip, gpd-corinthians-2013, shelden:2002,
schneider-aag12}.

\bigskip
\noindent\textbf{Contributions.} The main contributions of this paper are as follows:

\begin{itemize}
\item We present a novel optimization method for increasing the developability of an arbitrary surface. It is based on local approximations
of the surface by developable surfaces with planar and thus circular Gauss images. 
While we could also use other representations within our framework, we prefer B-splines in
order to have simple access to smoothness of patches. Moreover, we naturally obtain a patchwork 
of regular quad combinatorics, which is a preferred arrangement in many architectural projects. 

\item We provide a justification of our approach in two ways:  We discuss local approximations
of developable surfaces, especially with those being characterized by a planar Gauss image. 
Moreover, we study the implications of a nearly curve-like
Gauss image on the underlying surface, thus supporting our claim of achieving near developability
through Gauss image thinning. 

\item We introduce a variation of the main method presented in the paper to tackle the problem of paneling a freeform surface with (rotational) cylindrical, (rotational) conical and planar panels, which are the main preferred types of developable panels in architecture due to the reduced cost of manufacturing.

\item We provide results that illustrate the power of the proposed approach and outline
potential directions for future research.
\end{itemize}

\bigskip 
\noindent
\textbf{Overview of the paper.} This paper is organized as follows. In Section \ref{sec:fund}, we outline some 
important fundamentals for our work and, in section \ref{sec: method} present the main optimization algorithm step by step. Section \ref{sec: panelization} focuses on a variation of the main optimization algorithm which is
designed for paneling a freeform surface with panels that are special cases of developable surfaces. We present the differences with the main algorithm and introduce any necessary new tools. In section \ref{sec: experiments},
we provide results on various data sets, including ones from real architectural projects. Moreover, we discuss advantages and shortcomings of our approach and outline future work.


\section{Fundamentals}\label{sec:fund}

\subsection{Local approximations of developable surfaces}

We are interested in smooth or piecewise smooth developable
surfaces $S$. They are composed of $C^2$ surface patches which 
fall into one of the following four categories: planes, general cylinders,
general cones and tangent surfaces of space curves. 
Their {\it Gauss images} $C$, i.e. sets of unit normals viewed as points on the unit 
sphere $S^2$, are composed of {\it curves}. The junction points of $C$ where more than two curve
segments meet, correspond to 
planar patches on $S$. In the following, we discuss only the three non-trivial basic types:
These are ruled surfaces with a constant tangent plane along each ruling. In other words,
they are envelopes of a one-parameter family of planes. 

We are interested in second order local approximations of these basic types. 
 The following result is well-known (see, e.g. \cite[Theorem 6.1.4]{pottwall:2001}) and closely related to
the simple fact that the Gauss image of a developable surface is a spherical curve,
which has an osculating circle at each of its regular points. 

\begin{lemma} Along each ruling $r$, a non-planar developable ruled surface $S$ has second order
contact with a rotational cone $\Gamma$ (osculating cone). The vertex of this cone is the
singular point of $r$ (regression point). $\Gamma$ is a rotational cylinder for a 
cylindrical ruling $r$ (regression point at infinity) and it degenerates to a plane if $r$ is an inflection ruling.
\end{lemma}

Let us add a bit more detail for the generic case where $S$ is the tangent surface of a space
curve, $S:\ \xw(u,v)=\cw(u)+v\dot{\cw}(u)$. This so-called regression curve $\cw(u)$ is a singular curve on $S$. 
The osculating plane at $\cw(u)$, spanned by $\dot{\cw},\ddot{\cw}$, is the constant tangent plane of $S$ along a ruling 
(isoparameter line $u=const$). If $u$ is an arc length parameter, then the Frenet frame at $\cw(u)$
is given by the tangent vector $\ew_1=\dot{\cw}$, principal normal $\ew_2=\ddot{\cw}/\kappa$ (with curvature $\kappa=\|\ddot{\cw}\|$), 
and the binormal vector $\ew_3=\ew_1 \times \ew_2$. The Frenet equations can then be written in the
form $\dot{\ew}_i= \dw \times \ew_i$. Here $\dw=\tau \ew_1+\kappa \ew_3$ is the so-called Darboux vector, where $\tau$ denotes the
torsion. The Darboux vector is the direction vector of the osculating cone $\Gamma$. This means that the angle $\phi$ between
cone axis and ruling satisfies $\cot \phi = \tau / \kappa =:k$, a value which is called  {\it conical curvature} of the
developable surface at the ruling.

The Gauss image of a rotational cone $\Gamma$ is a circle $C$ on $S^2$ which becomes a 
great circle if $\Gamma$ is a cylinder and degenerates to a point for a plane $\Gamma$. So
all 2nd order local approximations addressed above have a {\it planar Gauss image} curve $C$.
However, a planar Gauss image $C$ of a surface $\Gamma$ does not yet imply that $\Gamma$ is a
cone, while $\Gamma$ must be a cylinder if $C$ is a great circle and a plane if
$C$ is just a point.  So let us discuss the case of a small circle $C$ as Gauss image of a surface.
These surfaces are well studied in classical differential geometry and known as {\it surfaces
of constant slope}. They are the tangent surfaces of curves $c$ of constant slope. Their tangents
form a constant angle with a certain direction in space, which is obviously the rotational axis
of the circle $C$. For a detailed study of these surfaces, we refer to \cite[Section 6.3]{pottwall:2001}. 
The increased degrees of freedom compared to the osculating cone allow us to increase the 
local approximation of an arbitrary developable surface by one with a planar Gauss image:

\begin{theorem} \label{thm1} At each regular point $p$ of a developable ruled surface $S$, there is a developable
surface $\Gamma$ with a planar Gauss image, which has 
second order contact with $S$ along the entire ruling through $p$ and interpolates a 
curve $a \subset S$ through $p$. 
\end{theorem}

\noindent {\em Proof.} We omit the cases where $S$ is a plane or a cylinder,
since these surfaces already have a planar Gauss image curve. So we are left with cones and tangent
surfaces $S$. We pick the osculating cone $\Gamma_p$ of $S$ along the ruling $r_p$ through $p$ and
intersect $S$ with the plane $A$ through $ p$ which is orthogonal to the axis of $\Gamma_p$. This 
yields the curve $a$. Note that the plane $A$ intersects the
cone $\Gamma_p$ in a circle, which is the osculating circle of $a$ at $p$. 
The construction of the
developable surface $\Gamma$ proceeds as follows: Through each tangent of $a$ we compute the two planes which form
the same angle
with the axis of $\Gamma_p$ as $\Gamma_p$ does. Among these two planes, we select the one which is closer to
the corresponding tangent plane of $S$. Then, the envelope of this family of planes is the desired developable
surface  $\Gamma$ with a planar Gauss image described in the theorem.  By construction, $\Gamma$ and $S$ share
the osculating cone $\Gamma_p$ and thus have second order contact along the ruling through $p$. We could choose 
another curve $a \subset S$
which lies transversal to the rulings of $S$, but leave it with this special choice as it simplifies the further 
analysis.\\


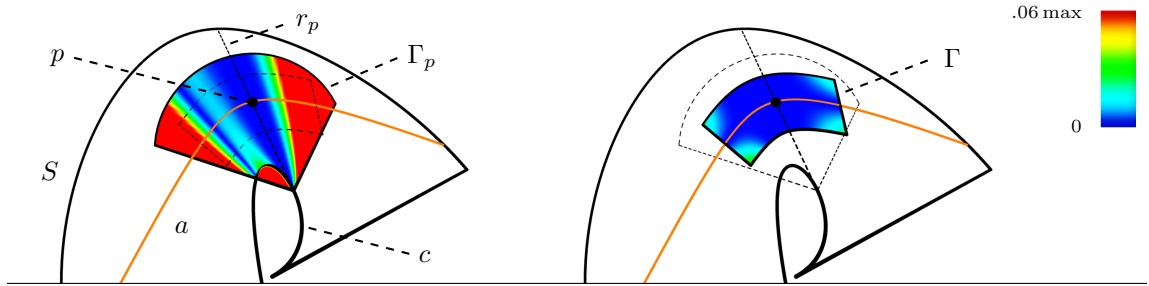
\begin{figure}[H]
	\centering
	\begin{tikzpicture}
	\node[anchor=south west,inner sep=0] (image) at (0,0) {\includegraphics[width=0.98\linewidth]{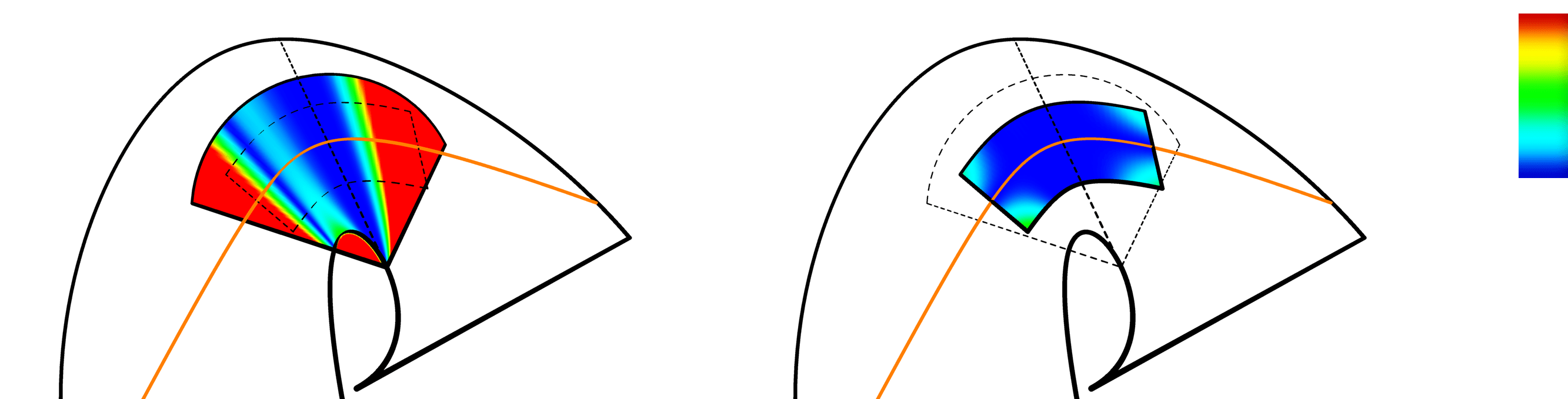}};
	\begin{scope}[x={(image.south east)},y={(image.north west)}]
	
	
	\node at (0.21, 0.64) {\textbullet};
	\draw[dashed,-,thick] (0.21, 0.64) -- (0.05,0.8);
	\node[left] at (0.05,0.8) {$p$};
	
	\node[left] at (0.045,0.4) {$S$};
	
	\node[right] at (0.132,0.2) {$a$};
	
	\draw[dashed,-,thick] (0.19, 0.85) -- (0.24,0.92);
	\node[right] at (0.24,0.92) {$r_p$};
	
	\draw[dashed,-,thick] (0.28, 0.7) -- (0.34,0.8);
	\node[right] at (0.34,0.8) {$\Gamma_p$};
	
	\draw[dashed,-,thick] (0.26, 0.2) -- (0.35,0.108);
	\node[right] at (0.35,0.10) {$c$};

	\node at (0.678, 0.64) {\textbullet};
	
	\draw[dashed,-,thick] (0.74, 0.68) -- (0.82,0.8);
	\node[right] at (0.82,0.8) {$\Gamma$};

	\node[left] at (0.96,0.56) {\scriptsize $0$};
	\node[left] at (0.96,0.96) {\scriptsize $.06\max$};
	
	
	\end{scope}
	\end{tikzpicture}%
	\vspace{-12pt}
	\par\noindent\hrulefill
	\caption{Local approximations of a developable surface $S$, which is the tangent surface of a space curve $c$. Left: The osculating cone $\Gamma_p$ at a point $p\in S$ approximates $S$ to 2nd order along the entire ruling 
	$r_p$. Right: A developable surface $\Gamma$ as in Theorem \ref{thm1} approximates $S$ even better,
	as is seen from the color coding of $\Gamma$ and $\Gamma_p$ according to their orthogonal
	distance to $S$.}
	\label{fig:planar_Gauss_image_approximation}
\end{figure}


For that, we use a local $(x,y,z)$ coordinate 
system with $A:z=0$ and describe the curve $a$ by its support function $h(u)$. This means that we
view $a$ as envelope of its tangent lines
$$ L(u):\  x \cos u + y \sin u + h(u) =0,$$
which form the angle $u$ with the $y$-axis and possess the signed distance $h(u)$ from the origin (if the
positive side of $L$ is determined by the normal vector $(\cos u, \sin u)$). The derivative with
respect to $u$ is the curve normal, $\dot{L}(u):\ - x \sin u + y \cos u + \dot{h}(u) =0$. Intersecting the two lines $L,\dot{L}$,
we obtain a parameterization of the curve $a$ as
$$  \aw(u):\ x=-h\cos u + \dot{h} \sin u, \ y=-h \sin u - \dot{h}\cos u. $$
Differentiating again yields the curvature centers (evolute) of $\aw(u)$ as $\aw^*(u)=\dot{L} \cap \ddot{L}$,
$$  \aw^*(u):\ x=\dot{h} \sin u+\ddot{h} \cos u, \ y= - \dot{h}\cos u + \ddot{h} \sin u. $$
Thus, the signed curvature radius of $\aw(u)$ is $\rho(u) = h(u) + \ddot{h}(u)$. 

Let $p$ be the point $\aw(0)=(-h(0), -\dot{h}(0),0)$. To shorten notation, we use the notation $h(0)=:h_0$ and likewise for the derivatives. Then the $z$-parallel line through the curvature center 
$\aw^*(0)=(\ddot{h}_0, -\dot{h}_0,0)$ is the axis of the osculating cone $\Gamma_p$. With $k$ as conical
curvature of $\Gamma_p$ and of $S$ at $u=0$, the vertex of $\Gamma_p$ has $z$-coordinate $z=(h_0+\ddot{h}_0)/k=\rho_0/k$. 
Planes $P(u)$ through the tangents of $\aw$ and with the same inclination against the $z$-axis as $\Gamma_p$ have the
equations
\begin{equation} P(u):\ x \cos u + y \sin u - kz+ h(u) =0. \label{const-slope}
\end{equation}
Their envelope is the desired approximation $\Gamma$ of $S$ at $p$ with a planar Gauss image and through $\aw$. 
Differentiating with respect to $u$ yields planes $\dot{P},\ddot{P}$ whose equations agree with those of 
$\dot{L},\ddot{L}$ and are therefore $z$-parallel planes through these lines. Recall that rulings of $\Gamma$ are obtained 
as intersections $P \cap \dot{P}$ and the regression curve is found as  $P \cap \dot{P} \cap \ddot{P}$. As discussed 
in more detail in \cite[Section 6.3]{pottwall:2001}, the regression curve of $\Gamma$ lies in the $z$-parallel cylinder 
through $\aw^*$ and the intersections of $\Gamma$ with planes $z=const$ are translated offsets of $\aw$. 
The intersection curve $\aw_1$ of $\Gamma$ with the plane $z=1$ is a translated version of the 
offset of $\aw$ at distance $k$ and therefore has a 
support function $h(u)-k$. The ruling vectors $\rw_1=\aw_1-\aw$ of $\Gamma$ are $\rw_1(u)=(k\cos u, k \sin u, 1)$. 

The intersection curve $\bar{\aw}$ of $S$ with $z=1$ has a support function $\bar{h}(u)=h(u)-k+f(u)$.
Due to the 2nd order contact at $u=0$, we have $f(0)=\dot{f}(0)=\ddot{f}(0)=0$. 
%
Then, the tangent planes of $S$ are
\begin{equation} T(u): \ x \cos u + y \sin u + (f(u)- k) z+ h(u) =0, \label{planes-S}
\end{equation}
and the ruling vectors of $S$ are $\rw=\bar{\aw}-\aw$,
$$ \rw(u)=((k-f)\cos u + \dot{f} \sin u, (k-f) \sin u - \dot{f} \cos u, 1).$$
%
%
%
Now we have parameterizations of $S$ as $\sw(u,v)=\aw(u)+ v \rw(u)$ and of $\Gamma$ as $\gw(u,v)=\aw(u)+v \rw_1(u)$,
which concludes the proof. 

However, we want to go beyond that and estimate the distance between $S$ and its approximation $\Gamma$,
and compare it to the distance between $S$ and the osculating cone $\Gamma_p$. 

We over-estimate the distances by
measuring them in planes $z=const=v$ and there between points with parallel tangents. This means that we
measure distances between points of the two surfaces which have the same parameter values $(u,v)$.
This distance $\delta (u,v)$ between $S$ and $\Gamma$ is given by
\begin{equation} \delta (u,v)=|v| \|\rw_1(u)-\rw(u)\|=|v| \sqrt{f(u)^2+\dot{f}(u)^2}. \end{equation}
We can also look at distances $\bar{\delta}$ between the parallel
tangents directly, which are in view of equations (\ref{const-slope}) and (\ref{planes-S}),
$$ \bar{\delta}(u,v)=|v f(u)|. $$
%
%
%
For $u=0$ we get the ruling $r_p$ through $p$ and of course $\delta,\bar\delta=0$.

%

Let us compare this with the approximation of $S$ by the osculating cone $\Gamma_p$. The cone
is given by (\ref{const-slope}) where $h$ is replaced by the support function $h_c$ of the osculating
circle $\cw_o$ of $\aw$ at $p=\aw(0)$, 
$$ h_c(u)=\rho_0+ \dot{h}_0\sin u-\ddot{h}_0\cos u. $$
The parameterization of the osculating circle is
$$ \cw_o(u)=(\ddot{h}_0 -\rho_0\cos u, - \dot{h}_0 -\rho_0\sin u, 0). $$
Thus, a parameterization of $\Gamma_p$ is given by $\cw_o(u)+v \rw_1(u)$, and the two errors 
$\delta_p, \bar{\delta}_p$ between $S$ and $\Gamma_p$ become
$$ \delta_p(u,v)=\|\cw_o(u)-\aw(u)+v(\rw_1(u)-\rw(u))\|,\  \bar{\delta}_p(u,v)=| vf(u)+h(u)-h_c(u)|.$$ 
To get better insight into the behavior of the errors, we insert Taylor expansions at $u=0$,
$$ f(u)=a_3u^3+\ldots,\  h(u)=h_0+\dot{h}_0u+\frac{\ddot{h}_0}{2}u^2+\frac{\dddot{h}_0}{3}u^3 +\dots.$$
The error vector between $\aw$ and $\cw_0$ now reads 
$$\cw_o(u)-\aw(u)=(-\frac{\dot{\rho}_0}{3} u^3+\ldots, \frac{\dot{\rho}_0}{2} u^2+\frac{\ddot{h}_0}{6} u^3+\ldots, 0).$$ 
Note that the quadratic term in the error vector  
is in tangential direction at $p$,
and thus confirms the 2nd order contact between $\cw_o(u)$ and $\aw(u)$ at $p$.  
For the errors, we find the following expansions,
$$ \delta(u,v)= |3a_3u^2v +\ldots|,\ \bar{\delta}(u,v)=|a_3u^3v+\ldots|,$$
and
$$ \delta_p(u,v)= |\frac{\dot{\rho}_0}{2} u^2+3a_3u^2v +\ldots|,\ \bar{\delta}_p(u,v)=|\frac{\dot{\rho}_0}{6} u^3+
 a_3u^3v+\ldots|.$$
As expected, the approximation of $S$ by the osculating cone $\Gamma_p$ is not as good as with $\Gamma$,
since the deviation in the base plane $z=0\ (v=0)$ adds to the error everywhere. The appearance of 
the derivative $\dot{\rho}_0$ of the curvature radius $\rho(u)$ at $u=0$ in the lowest order term is no surprise, as
for $\dot{\rho}_0=0$ the osculating circle $\cw_o$ has 3rd order contact with $\aw$ and $S$ at $p$. 

There is one exception which we did not cover here, namely if the ruling $r_p$ through $p$ is an {\it inflection
ruling}. In that case, $\Gamma_p$ degenerates to the tangent plane, and one cannot parameterize directly via the
tangent directional angle $u$. Instead, one can use another parameter $t$, and work with a parameterization in
support coordinates $(u(t),h(t))$, as in \cite[pp. 362-363]{pottwall:2001}.

Knowing that surfaces with a planar Gauss image approximate developable surfaces
at each point so well, we can increase developability by enforcing local
approximations of this type through an optimization algorithm (see section \ref{sec: method}). 

\subsection{Surfaces with a thin Gauss image}\label{subsec:thin}

Our method will try to make the Gauss image of a B-spline surface
thinner. After that, it will lie in a region $R_\varepsilon$ on the sphere which 
has at most geodesic distance $\varepsilon$ to a 
curve $C \subset S^2$. Let us briefly discuss the implications on
a surface $S$ which has a Gauss image in such an $\varepsilon$-strip $R_\varepsilon$.
For that, we pick a part of the surface without an umbilic; there the
principal curvature lines form a quadrilateral curve network without
singularities. For simplicity, let us just consider a patch $\mathcal{P} \subset S$ in this
region which is bounded by four principal curvature lines and does not contain parabolic points.
Moreover, we select a square-like patch $\mathcal{P}$, meaning that the average length of the two pairs of opposite
boundary curves is the same. 
The Gauss image $\sigma(\mathcal{P})$ of that principal patch $\mathcal{P}$ is a principal patch on $S$; corresponding
curves on $P$ and $\sigma(\mathcal{P})$ have parallel tangents at corresponding points, as they are principal directions
and thus eigendirections of the derivative of the Gauss map. As we exclude parabolic points in $\mathcal{P}$, the
Gauss map is regular everywhere and thus locally injective. 

The Gauss image $\sigma(\mathcal{P})$ of $\mathcal{P}$ is squeezed into the thin region $R_\varepsilon$.  Being contained in $R_\varepsilon$,
at least one family $F_1$ of principal curvature lines on $\mathcal{P}$ must be mapped to very short curves in $R_\varepsilon$.
If this is not true for the other family $F_2$ of principal curvature lines; the Gauss image curves of that
family must be nearly parallel to the central curve $C$ of $R_\varepsilon$. Thus, the Gauss images of curves
in $F_1$ will be nearly orthogonal to $C$ (see Figures \ref{fig:principal}, \ref{fig: principal net}). 
Their length can be bounded depending on the width variation of $\sigma(\mathcal{P})$. The shortening of curves in 
$F_1$ through the Gauss map to a length $\approx \varepsilon$
implies that the curves themselves will be close to straight lines. A surface with one family of 
straight principal curvature lines is exactly developable; our surface is only an approximation of that. 
A more thorough investigation
of the geometric implications of a thin Gauss is left for future research.\\


\begin{figure}[H]
	\centering
	\begin{tikzpicture}
	\node[anchor=south west,inner sep=0] (image) at (0,0) {\includegraphics[width=0.98\linewidth]{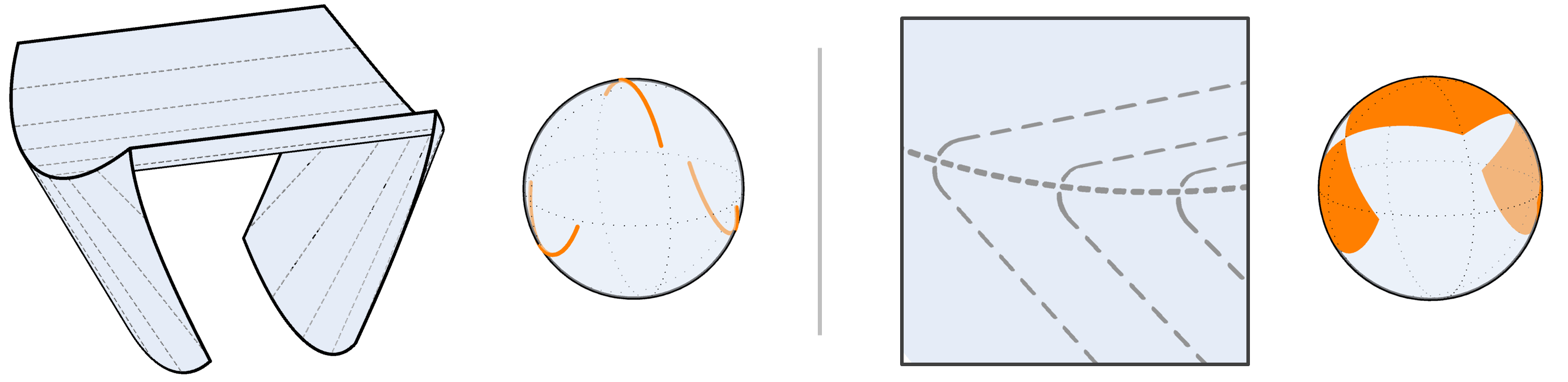}};
	\begin{scope}[x={(image.south east)},y={(image.north west)}]
	
	\node[above right] at (0,0) {\small\textbf{(a)}};
	\node[above left] at (1,0) {\small\textbf{(b)}};
	
	\node[above right] at (0.26,0.06) {$S$};
	\node[above] at (0.405,0.85) {$\sigma(S)$};
	
	\node[above right] at (0.58,0.06) {$S^\prime$};
	\node[above] at (0.91,0.85) {$\sigma(S^\prime)$};

	\draw[dashed,-,thick] (0.01, 0.8) -- (0.3,0.95);
	\node[right] at (0.3,0.95) {$f_1$};
	\draw[dashed,-,thick] (0.26, 0.75) -- (0.3,0.8);
	\node[right] at (0.3,0.8) {$f_2$};

	\draw[dashed,-,thick] (0.36, 0.34) -- (0.36,0.2);
	\node[below] at (0.36,0.2) {$C_1$};
	\draw[dashed,-,thick] (0.42, 0.7) -- (0.47,0.8);
	\node[right] at (0.47,0.8) {$C_2$};
	\draw[dashed,-,thick] (0.45, 0.44) -- (0.45,0.2);
	\node[below] at (0.45,0.2) {$C_3$};
	
	\end{scope}
	\end{tikzpicture}%
	\par\noindent\hrulefill
	\caption{\textbf{(a)} Example of a developable shape $S$ with curved folds $f_1$, $f_2$, and its Gauss image $\sigma(S)=C_1\cup C_2\cup C_3$. \textbf{(b)} Rounding the fold curves of $S$, leads to shape $S^\prime$ with Gauss image $\sigma(S^\prime)$ which is not thin.}
	\label{fig: fat gauss}
\end{figure}
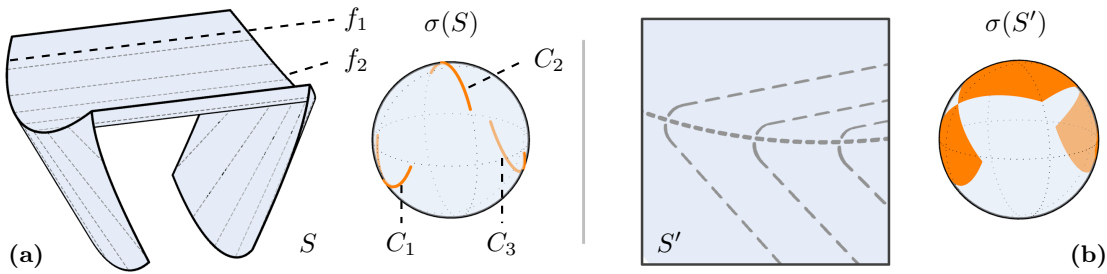


Due to our focus on architectural geometry, we can exclude surfaces with wrinkles or folds appearing for example
in cloth. These wrinkles are close to curves formed by parabolic points and have one very high principal curvature.
They are not of interest in the present paper, and are not characterized by thin Gauss images. Some insight into the geometry
of these folds can be obtained as follows: Consider a planar sheet of material, mark a fold curve on it and bend it into a 3D shape $S$,
leading to a developable surface with a curved crease (for the local geometry of such curved folds, see e.g. (\cite[Section 6.5]{pottwall:2001}). The two developable surfaces on either side of the fold curve $f$ have
curves $C_1, C_2$ as Gauss images.  Now let us add a thin smooth blend to round off the fold curve $f$. The Gauss image of that blend surface
will connect the two curves $C_1, C_2$ to a region which needs not be thin at all. With a sufficiently small blending radius
the shape $S$ can be arbitrarily close to
an exact developable surface and thus be nearly developable, but the Gauss image will not be thin (see Figure \ref{fig: fat gauss}).

Therefore, our
approach of thinning the Gauss image implies the construction of nearly developable surfaces, but the converse
is not true. A nearly developable surface needs not have a thin Gauss image, due to the phenomenon of
wrinkles. For materials which allow only very little stretching, these wrinkles appear to be smoothed versions
of developable surfaces with curved folds, as indicated above. There is interesting research on this phenomenon,
combining geometry and physics; see e.g. \cite{Cerda}.  However, we are not aware of any differential geometric
characterization of nearly developable surfaces which does not use the planar unfolding.

\subsection{Developable bicubic surfaces}

We will use bicubic B-spline surfaces and thus it is appropriate to justify
this choice. When it comes to modeling nearly developable surfaces, our choice is natural due to
the approximation power of splines. The condition of one family of nearly straight principal curvature
lines is sufficiently soft to be modeled nicely with these splines.

However, especially in our architectural application, we will model panel arrangements also by bicubic B-spline surfaces,
with knots of multiplicity three, which are just $C^0$ patchworks of bicubic polynomial
patches. We want these polynomial patches to be close to developable surfaces, in particular to 
right circular cones or cylinders. Thus, we briefly discuss {\it developable bicubic surfaces}. \\

\noindent {\bf Bicubic patches on tangent surfaces.} The tangent surface of a polynomial
cubic $\cw(u)$ can be parameterized as 
$$\xw(u,v)=\cw(u)+v\dot{\cw}(u),$$
 and it is therefore a bicubic
surface. In this form, the rulings are $v$-isoparameter curves and an axis aligned rectangle in the
parameter domain represents a patch on the surface bounded by two rulings. There are other bicubic
patches on that surface, which are obtained as images of arbitrary parallelograms in the $(u,v)$-plane.
Equivalently, one can obtain them as images of the unit square $[0,1]^2$ in a $(\bar{u},\bar{v})$ parameter
plane via an affine parameter change,
$$ u=a_0+a_1\bar{u}+a_2\bar{v},\ v=b_0+b_1\bar{u}+b_2\bar{v}.$$
Furthermore, special bilinear re-parameterizations where the first equation remains and the second one reads 
$$  v=b_0+b_1\bar{u}+b_2\bar{v}+b_3\bar{u}\bar{v},$$
also yield bicubic patches on that tangent surface. 

Even {\it the tangent surface of a polynomial quartic $\cw(u)$ has a bicubic parameterization}. We
write $\cw=\aw_4u^4+\aw_3u^3+\ldots+\aw_0$ in monomial form and parameterize its tangent surface as
$$\xw(u,v)=\cw(u)+(-u/4+v)\dot{\cw}(u),$$
which is a bicubic representation. A complete classification of all bicubic tangent surfaces
is an open problem. For our purposes it suffices to see that tangent surfaces of quartic curves 
are included in this class of surfaces, which leaves sufficient flexibility for modeling.\\

\noindent {\bf Bicubic patches on cones and cylinders.}
A cone with vertex $\vw$ can be written as $\xw(u,v)=\vw+f(u,v)\cw(u)$. To get a bicubic parameterization,
we can use a cubic curve $\cw(u)$ and a cubic polynomial $f(u,v)=g(v)$  or a quadratic curve (parabola) $\cw(u)$ and a function $f(u,v)$ of bi-degree $(1,3)$. In the former case, the cone is in general a cubic surface, while in the
latter case one parameterizes quadratic cones. 

A cylinder $\xw(u,v)=\aw(u)+f(u,v)\rw$, with a ruling direction $\rw$, has a bicubic representation when
$\aw(u)$ is at most cubic and $f$ any bicubic function.\\

\noindent{\bf Developable bicubic patches with a planar Gauss image.} This class of surfaces includes all
bicubic cylinders. Among the cones, only rotational cones are possible. We can generate them from the
special cone $x^2+y^2=z^2$, and then apply uniform scaling in $z$-direction and a rigid body motion. The special
cone is parameterized by a Pythagorean triple of bicubic functions $x(u,v),y(u,v),z(u,v)$ of the
form
$$ x(u,v)=2abw,\ y(u,v)=(a^2-b^2)w,\  z(u,v)=(a^2+b^2)w, $$
where $a(u,v)$, $b(u,v)$, $w(u,v)$ are bilinear functions. Bicubic tangent surfaces with a planar Gauss image
have a regression curve $\cw(u)$ of constant slope. It follows from our considerations above that the tangent
surface of
a polynomial curve $\cw(u)$ of constant slope and degree $\le 4$ is such a surface. These curves
 $\cw(u)$ are exactly the {\it spatial Pythagorean hodograph curves} of degree $\le 4$. For their generation and 
degrees of freedom, we point to the
monograph by R.~Farouki \cite[Chapter 21]{farouki:2008}. 

We have already mentioned rotational cones and note that rotational cylinders do not possess an
exact bicubic parameterization. This is due to the fact that a rotational cylinder cannot carry
a polynomial curve transversal to the rulings as it would project onto a circle. While a circle
does not have an exact polynomial parameterization, it is possible to achieve good
approximations with cubics (see \cite{zagar:2018} and the references therein). This is sufficient for our purposes.\\

\noindent {\bf Developable B-spline surfaces.} If two algebraic developable surface
patches
meet with $C^1$ continuity at a common curve (different from a ruling), their set of tangent planes 
agrees there. Due to the algebraic nature, agreement of the set of tangent planes along a curve segment 
is sufficient for
the agreement of the set of tangent planes everywhere and for agreement of the two algebraic surfaces. 
Therefore, any developable B-spline surface with
$C^1$ continuity represents a single polynomial developable surface, unless the patches are joined along
rulings. This latter case is used in \cite{chengcheng-dev-2015}. The former case is
useful to represent appropriate trimmed patches on polynomial developable surfaces, but not for
increasing the flexibility in modeling the surfaces themselves.

A regular bicubic surface $\Sw$ parameterized by parameters $u$, $v$ is developable when the Gaussian curvature vanishes at every point $(u,v)\in D$ of the surface. Based on this definition of developable surfaces, we can compute the algebraic complexity of the developability property for $\Sw$. Since the Gaussian curvature is the ratio of the determinants of the second and first fundamental forms, it is sufficient for the following equation to hold
$$\det (\mathbf{II}) = 0 \Leftrightarrow [\Sw_{uu}, \Sw_u, \Sw_v] [\Sw_{vv}, \Sw_u, \Sw_v] - [\Sw_{uv}, \Sw_u, \Sw_v]^2 = 0,~~ \forall (u,v)\in D $$
where $[\aw , \bw , \cw ]$ denotes the triple product of vectors $\aw$, $\bw$, $\cw\in \mathbb{R}^3$. Expanding and grouping with respect to monomials in parameters $u$, $v$ we get a polynomial  $f\in\mathbb{R}[x_{00},y_{00},z_{00},\ldots ,x_{33},y_{33},z_{33}] [u,v]$, where $(x_{ij}, y_{ij}, z_{ij})\in \mathbb{R}^3$, are the coordinates of control point $\Pw_{i,j}$ of surface $\Sw$. Following this grouping, we count that polynomial $f$ has 191 coefficients  $g_k\in\mathbb{R}[x_{00},y_{00},z_{00},\ldots ,x_{33},y_{33},z_{33}]$, where $k=1,\ldots,191$.

The requirement that polynomial $f$ vanishes for all values $(u,v)\in D$ is satisfied if $f$ is identically the zero polynomial, or equivalently all coefficient polynomials $g_k$ vanish. This means that, if we need to guarantee these conditions precisely by evaluating $f$ at different points on the surface, we would require a minimum of 191 points in a general position, namely points that would generate linearly independent combinations of $g_k$ . In practice, since $\deg_u(f)=\deg_v(f)=13$ we would define a $14\times 14$ regular grid over $D$ to acquire 196 evaluation points.

Alternatively, we can examine the algebraic variety $V(I)$ of the ideal $I=\left\langle g_1,\ldots ,g_{191}\right\rangle$ generated by the coefficient polynomials $g_k$. Again, these are 191 homogeneous polynomials in 48 variables with $\deg(g_k)=6$. Computing a reduced Gr\"obner basis in an attempt to work with a minimal number of generators $h_m\in\mathbb{R}[x_{00},y_{00},z_{00},\ldots ,x_{33},y_{33},z_{33}]$, with $m\leq 191$, for the ideal $I$ is computationally expensive, and is expected to produce generators that have increasingly higher degrees \cite{Dube:1990:SPI:102351.102365}.

These observations only demonstrate that if we wish to increase interactivity in the design process with developable surfaces, we need to avoid the computational complexity of exact satisfiability and instead sufficiently approximate the developability property in an efficient way.


\section{Increasing developability}\label{sec: method}


Motivated by Theorem \ref{thm1}, we can try to increase the developability of a surface $S$ by ensuring
 that the Gauss images of well chosen regions on $S$ are nearly planar.  
%
%
%
Using this basic idea, we now discuss
the details of an optimization algorithm which iteratively deforms a bicubic B-spline surface towards
a nearly developable one.

\subsection{Optimization setup}\label{sec: optimization setup}

\smallskip\noindent\textbf{Surface.} Let us consider a bicubic B-spline surface $\mathbf{S}: \mathbb{R}^2\rightarrow\mathbb{R}^3$, 
\begin{equation}\label{eq:surface}
\mathbf{S}(u,v)=\sum\limits_{i=0}^{n}\sum\limits_{j=0}^m B_{i,3}(u) B_{j,3}(v) \mathbf{P}_{i,j},
\end{equation}
where $u,v\in [0,1]$ and $B_{i,3}(u)$, $B_{j,3}(v)$ are cubic B-spline basis functions defined on uniform knot sequences in both directions. $\{\mathbf{P}_{i,j}\}\in \mathbb{R}^3$ are the control points of the surface $\mathbf{S}$, where $0\leq i\leq n$, $0\leq i\leq m$ and $n,m\geq 3$. For more information on B-spline surfaces and NURBS surfaces in general, we direct the reader to \cite[Section 4.4]{Piegl:1997:NB:265261}.

Surface $\mathbf{S}$ serves as the central object of study in this work. A generic surface of the above form is non-developable and we aim to increase its developability by modifying the coordinates of its control points in a "minimal" way that will be defined in the following sections.

We point out that surface $\mathbf{S}$ could be defined as any NURBS surface as long as the weights of the control points and the knot vectors are fixed and are not considered variables in the optimization process. This simplifies and accelerates the optimization procedure while not sacrificing the quality of our results in the sense that B-spline surfaces are adequate approximations of more general NURBS surfaces.  For readability, we define $\mathbf{S}$ as an elementary B-spline surface while keeping in mind that the following applies to more general surfaces.\\

\noindent\textbf{Sampling the surface.} We begin by sampling $\mathbf{S}$, the surface that is to be optimized, at a set of evaluation points $\{\mathbf{p}_k\}\subset \mathbb{R}^3$, which we will call \emph{sample points}.

The approach we took for the sampling was to uniformly sample the parameter space, motivated by the fact that convoluted areas on the surface $\mathbf{S}$, i.e. areas where the control points are concentrated and finer features emerge, would be represented by more evaluation points inherently. We set the number of sample points $L_u$, $L_v$ along the $u$, $v$ directions respectively and get a gridded pattern of points $(u,v) \in [0,\frac{1}{L_u + 1},\ldots ,1]\times [0,\frac{1}{L_v + 1},\ldots ,1]$ on the parameter space, which in turn results in the set of required sample points $\{\mathbf{p}_k\}$ on the surface $\mathbf{S}$.

The evaluation of points $\mathbf{p}_k$ is given by formula \ref{eq:surface}, which is linear in the coordinates of the control points with constant coefficients. In practice, these coefficients are precomputed per point and stored. Whenever the control points are updated by the optimization process or user input, we re-evaluate the position of the sample points using the stored coefficients.\\

\noindent\textbf{Grouping into patches.} Next, we consider overlapping neighborhoods on the surface, that we will call \emph{patches}, and that are represented as sets of sample points $U_j$. We construct the patches in such a way that neighboring patches will have non-empty intersections, i.e. there exists at least one sample point that belongs to both patches. The importance of this property will become clear in a later section.

\noindent\hrulefill
\vspace{-6pt}
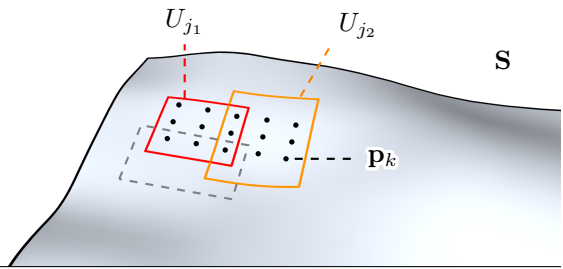
\begin{figure}[H]
\floatbox[{\capbeside\thisfloatsetup{capbesideposition={left,bottom},capbesidewidth=0.45\linewidth}}]{figure}[\linewidth]
{\caption{Surface $\mathbf{S}$ is sampled at various evaluation points $\mathbf{p}_k$. The sample points are then grouped to overlapping groups. An example of such a grouping are groups $U_{j_1}$ and $U_{j_2}$.\\~\\~\\}\label{fig:patches}}
{\begin{tikzpicture}
	\node[anchor=south west,inner sep=0] (image) at (0,0) {\hspace{8pt}\includegraphics[width=\linewidth]{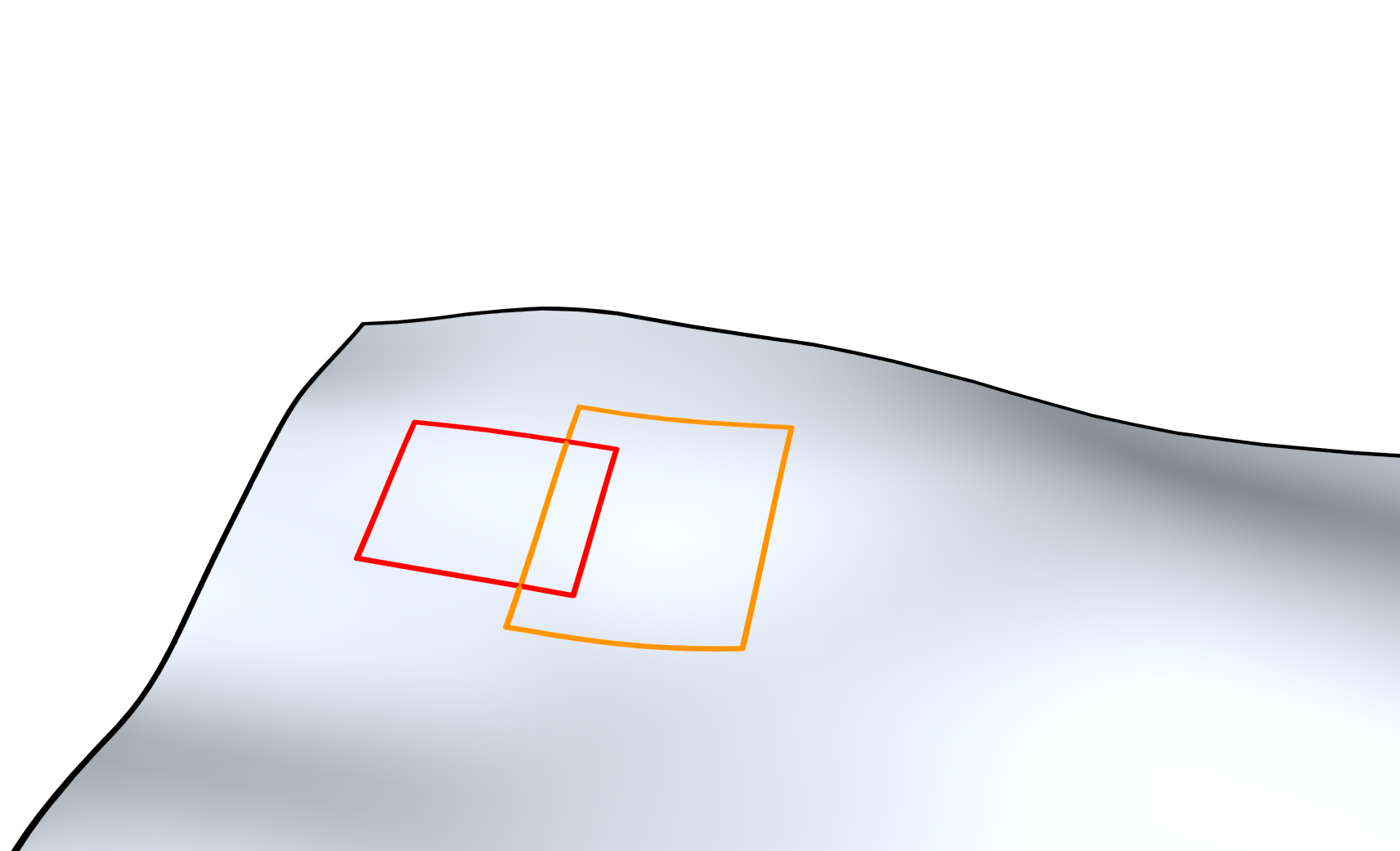}};
	\begin{scope}[x={(image.south east)},y={(image.north west)}]
    
    \foreach \Point in {(0.44,0.446), (0.43,0.396), (0.42,0.346), (0.39,0.465), (0.38,0.415), (0.37,0.365), (0.34,0.48), (0.33,0.43), (0.32,0.38), (0.492,0.435), (0.485,0.385), (0.477,0.335), (0.542,0.42), (0.535,0.37), (0.525,0.32)}{
              \fill \Point circle[radius=1pt];
          }
    
	\node[above] at (0.9,0.56) {$\mathbf{S}$};
    
	\draw[red, dashed,-, thick](0.35,0.5) -- (0.35,0.66);
    \node[above] at (0.35,0.66) {$U_{j_1}$};
    
    \draw[orange, dashed,-, thick](0.55,0.505) -- (0.6,0.65);
    \node[above right] at (0.6,0.65) {$U_{j_2}$};
    
    \draw[dashed,-, thick](0.54,0.32) -- (0.65,0.32);
    \node[right] at (0.65,0.32) {\contour{white}{$\mathbf{p}_k$}};
    
    \draw[dashed,gray,-, thick] (0.237,0.26) -- (0.275,0.42) -- (0.46,0.36) -- (0.432,0.2) --  (0.237,0.26);

    \end{scope}
\end{tikzpicture}}
\end{figure}
\vspace{-23.6pt}
\par\noindent\hrulefill\\


By uniformly sampling the parameter space we also simplify the process of grouping the sample points. The patches on the surface, as already mentioned, are represented by sets of sample points. By using the grid of points on the parameter space we can determine the patches just by setting the number of sample points in each of the $u$, $v$ directions that a patch will contain and the number of sample points that will belong in the overlap region for each of the $u$, $v$ directions. Figure \ref{fig:patches} focuses on two such patches as an example of a simple grouping.\\

\noindent\textbf{Normal computation.} We associate each sample point $\mathbf{p}_k$ with the unit normal $\mathbf{n}_k$ of the surface at that point. The unit normals define the \emph{Gauss map} $\sigma$ of the surface.
We compute the unit normal $\mathbf{n}_k$ of the surface point $\mathbf{p}_k$ as
$$
\mathbf{n}_k \coloneqq \sigma(\mathbf{p}_k) = \frac{\mathbf{S}_u \times \mathbf{S}_v}{\| \mathbf{S}_u \times \mathbf{S}_v \|},
$$
where $\mathbf{S}_u$, $\mathbf{S}_v$ are the partial derivatives of $\mathbf{S}$ with respect to $u$ and $v$. 
Note that $\mathbf{S}_u$ and $\mathbf{S}_v$,
$$
\mathbf{S}_u(u,v)=\sum\limits_{i=0}^{n}\sum\limits_{j=0}^m B_{i,3}^{(1)}(u) B_{j,3}(v) \mathbf{P}_{i,j}, \ 
\mathbf{S}_v(u,v)=\sum\limits_{i=0}^{n}\sum\limits_{j=0}^m B_{i,3}(u) B_{j,3}^{(1)}(v) \mathbf{P}_{i,j},
$$
are linear combinations of the control points with coefficients which we precompute and store to accelerate future computations \cite[Section 1.5]{Piegl:1997:NB:265261}.\\

\noindent\textbf{Gauss map of a patch.} For every patch $U_j$, we denote by $N_j$ the Gauss image of $U_j$, i.e. the set of unit normals $\mathbf{n}_k$ corresponding to the sample points $\mathbf{p}_k\in U_j$, 
$$
N_j = \sigma(U_j) = \sigma(\{ \mathbf{p}_k \}) = \{ \mathbf{n}_k \}.
$$
We associate each patch $U_j$ with a plane $H_j\subset \mathbb{R}^3$ with equation  $\mathbf{v}_j \cdot \mathbf{x} + d_j = 0$. Here, $\mathbf{v}_j$ is a unit normal vector of $H_j$ and $d_j$ is the distance of $H_j$ from the origin. $H_j$  serves as the target plane for $N_j$. By optimization, we will enforce all normal vectors in $N_j$ to lie on $H_j$ and thus aim at a planar Gauss image of patch $U_j$.\\

\subsection{Initialization}\label{sec: initialization}

The variables of the optimization are the coordinates of the control points $\mathbf{P}_{i,j}$ and the cutting planes $H_j$ that define the Gauss image circles per patch $U_j$. In this section, we describe the initialization step of the optimization process.\\

\noindent\textbf{Control points.} We assume that we always have an initial state for the surface that is either user defined or is provided by other means. We initialize the control point coordinates with the values from this initial configuration. Those in turn will be used to initialize $H_j$ for every patch.\\

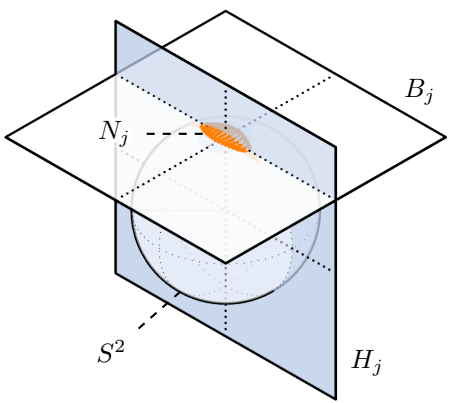
\begin{figure}[H]
\floatbox[{\capbeside\thisfloatsetup{capbesideposition={right,bottom},capbesidewidth=0.4\linewidth}}]{figure}[\linewidth]
{\caption{Consider the Gauss image $N_j$ of a group $U_j$. Plane $B_j$ is the best fitting plane to $N_j$, in the sense that it minimizes the sum of squared distances of points $N_j$ to the plane, and is considered an undesired initialization. Using $B_j$ as a target plane for the points in $N_j$ will degenerate the Gauss image to a single point, meaning patch $U_j$ will be flat. Alternatively, plane $H_j$ is the resulting plane from optimization problem \ref{op: plane initialization optimization problem} and captures the overall main principal direction of patch $U_j$. Plane $H_j$ is a better initial target plane, since it will not necessarily lead to a 0-dimensional Gauss image.}\label{fig: plane initialization}}
{\hspace{-2cm}
\begin{tikzpicture}
	\node[anchor=south west,inner sep=0] (image) at (0,0) {\includegraphics[width=0.8\linewidth]{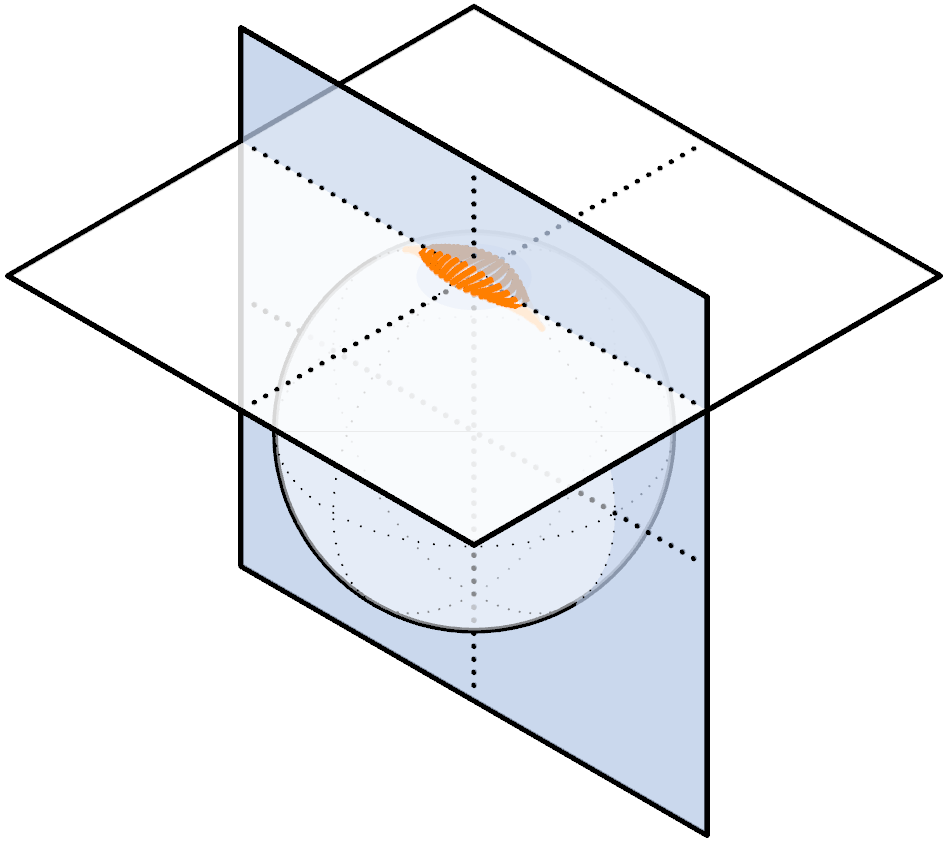}};
	\begin{scope}[x={(image.south east)},y={(image.north west)}]
    
	\node[above right] at (0.88,0.73) {$B_j$};
    
	\draw[dashed,-, thick](0.45,0.68) -- (0.31,0.68);
    \node[left] at (0.31,0.68) {\contour{white}{$N_j$}};
    
    \draw[dashed,-, thick](0.4,0.28) -- (0.3,0.18);
    \node[below left] at (0.3,0.18) {$S^2$};
    
    \node[right] at ((0.76,0.1) {$H_j$};
    
    
    \end{scope}
\end{tikzpicture}}
\end{figure}

\noindent\textbf{Cutting planes.} We want to optimize for planarity of the Gauss image $N_j$ of each patch $U_j$
and thus associate with each patch $U_j$ a target plane $H_j$ for $N_j$. 
Initializing the target plane $H_j$ for each patch with the best fitting plane to points $\mathbf{n}_k \in S^2$ works in the case that $U_j$ is a developable patch. However, this method does not produce the desired results if the patch is non-developable, as seen in Figure \ref{fig: plane initialization}. To overcome this, we use the following approach.\\


\noindent\hrulefill
\vspace{5pt}
\noindent\hrulefill\\


Consider the main principal direction $\mathbf{q}_k\in \mathbb{R}^3$ of surface $\mathbf{S}$ at point $\mathbf{p}_k$, i.e. the principal direction corresponding to the principal curvature with the maximum absolute value, that is $\max\{|\kappa_1(\mathbf{p}_k)|, |\kappa_2(\mathbf{p}_k)|\}$ where $\kappa_i:\mathbf{S}\rightarrow \mathbb{R}$, $i=1,2$, are the principal curvatures of a point on $\mathbf{S}$. The \emph{principal curvatures} and \emph{principal directions} of a surface at a point on the surface are the eigenvalues and corresponding eigenvectors of the \emph{shape operator} $-d_v\mathbf{N}=-\mathbf{I}^{-1}\mathbf{II}$, where $\mathbf{I}$, $\mathbf{II}$ are the \emph{first} and \emph{second fundamental forms} of the surface. We denote by $Q_j$ the set of main principal directions $\mathbf{q}_k$ corresponding to the points $\mathbf{p}_k\in U_j$.

We initialize $H_j$ as the plane passing through the barycenter of $N_j$ with unit normal in the direction of the vector which is "as orthogonal as possible" to the set $Q_j$ of main principal directions. Intuitively, we wish the initial cutting plane to intersect the sphere at a circle whose tangent at every point $\mathbf{c}\in S^2\cap H_j$ is "as parallel as possible" to the main principal directions of the sample points corresponding to the unit normals around $\mathbf{c}$.

In this way, the cutting plane serves as a generalized main principal plane, or a plane containing the main principal directions of every sample point in the patch. For a patch that is non-developable, we wish to initialize this main principal plane by using the main principal directions of the sample points weighted by a measure of confidence. A low
weight indicates the difficulty in distinguishing between the two principal curvatures. 
Specifically, we introduce weight $w_k\in [0,1]$ corresponding to each sample point $\mathbf{p}_k$ as
\begin{equation}\label{eq: weight}
w_k = 1 - \frac{\min \{|\kappa_i(\mathbf{p}_k)|\}}{\max \{|\kappa_i(\mathbf{p}_k)|\}},\quad i=1,2
\end{equation}
Now, for each patch $U_j$ we need to solve the following optimization problem. 


\begin{algorithm}[H]
\caption{Plane initialization}
\label{op: plane initialization optimization problem}
\mathleft
\begin{equation*}
\begin{aligned}
& \text{minimize}&& \sum\limits_{\mathbf{q}_k\in Q_j} w_k (\mathbf{v}_j\cdot \mathbf{q}_k)^2 \\
& \text{subject to}\quad && \mathbf{v}_j^2=1
\end{aligned}
\end{equation*}
\mathcenter
\end{algorithm}


Optimization problem \ref{op: plane initialization optimization problem} is a special case of minimizing a quadratic form under a quadratic regularization constraint.  Bringing the objective function into the form $\mathbf{v}_j^\top \mathbf{Q}\mathbf{v}_j$, the minimizer $\mathbf{v}_j^*$ is the normalized eigenvector corresponding to the smallest eigenvalue of $\mathbf{Q}$. Then, plane $H_j$ is given by 
$\mathbf{v}_j^* \cdot \mathbf{x} + d_j = 0$, with
$$
d_j = -\mathbf{v}_j^*\cdot \frac{1}{|N_j|}\sum\limits_{\mathbf{n_k}\in N_j}\mathbf{n}_k,
$$
where $|N_j|$ is the cardinality of $N_j$.

\subsection{Problem formulation}\label{subsec: problem formulation}


\noindent\hrulefill
\vspace{6pt}
\begin{figure}[H]
\floatbox[{\capbeside\thisfloatsetup{capbesideposition={right,bottom},capbesidewidth=0.4\linewidth}}]{figure}[\linewidth]
{\caption{The Gauss image $N_j$ of a single non-developable patch $U_j$ is a 2-dimensional subset of $S^2$. The cutting plane $H_j$ serves as the target plane for the normals $\mathbf{n}_k\in N_j$.\\~}\label{fig: patch gauss image}}
{\hspace{-30pt}
\begin{tikzpicture}
	\node[anchor=south west,inner sep=0] (image) at (0,0) {\includegraphics[width=0.9\linewidth]{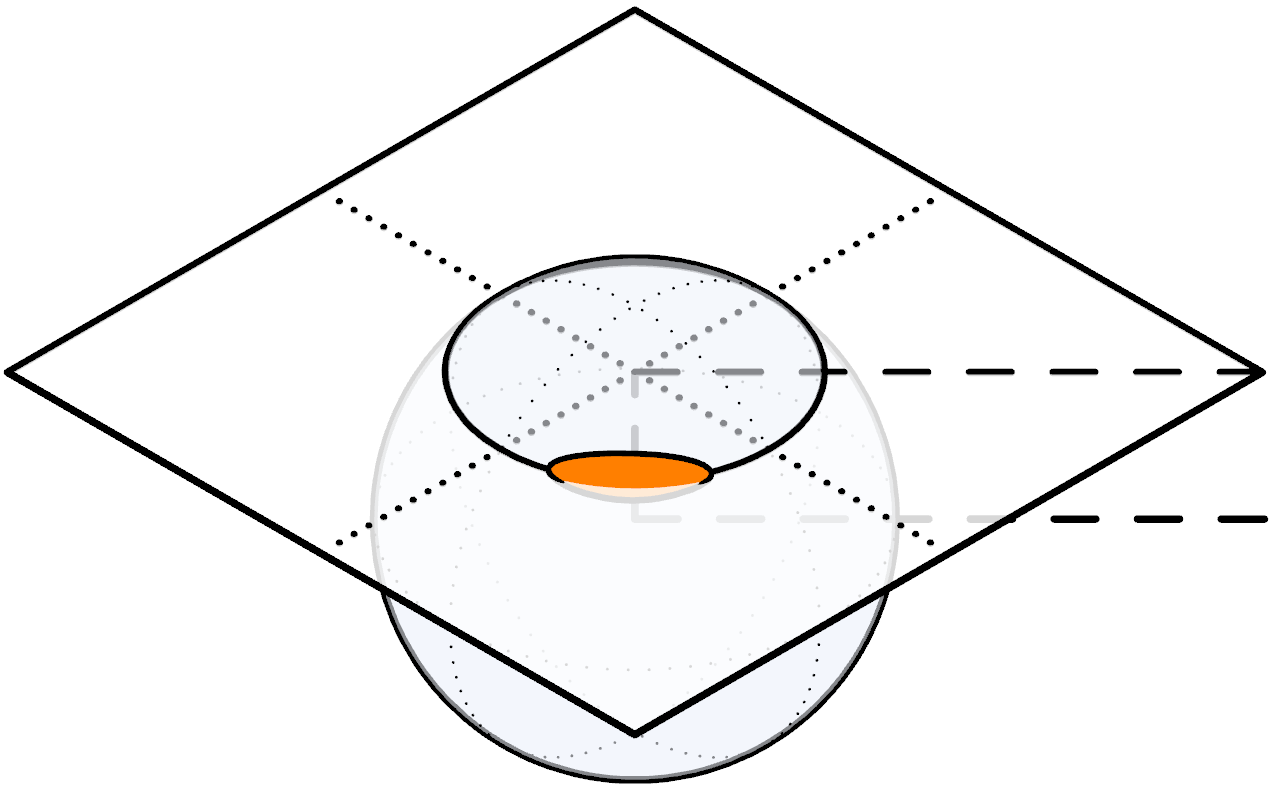}};
	\begin{scope}[x={(image.south east)},y={(image.north west)}]
    
	\node[above left] at (0.1,0.62) {$H_j$};
    \node at (0.99,0.43) {$d_j$};
    \node[below left] at (0.32,0.1) {$S^2$};
    
	\draw[dashed,-, thick](0.5,0.39) -- (0.78,0.12);
    \node[below right] at (0.78,0.12) {$N_j$};
    
    \draw[->,very thick, >=latex](0.5,0.53) -- (0.5,0.79);
    \node[right] at ((0.5,0.79) {$\mathbf{v}_j$};
    
    \end{scope}
\end{tikzpicture}}
\end{figure}
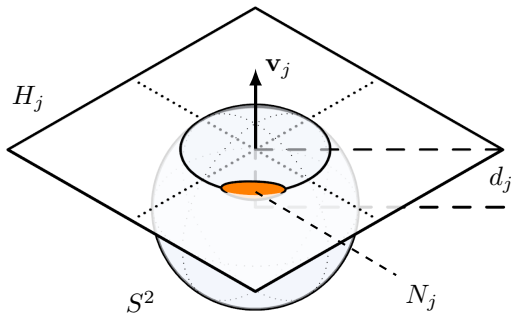
\vspace{-6pt}
\noindent\hrulefill\\


\noindent\textbf{Developability energy.} We are now ready to formulate the desired property of each patch to have a planar Gauss image by introducing an appropriate energy term $E_{\text{d}}$. This energy term measures per patch the total sum of distances of the normals  $\mathbf{n}_k \in S^2$ to the target patch plane, that is the quantity
\begin{equation}\label{eq: developability energy term before}
\sum\limits_j\sum\limits_{\mathbf{n}_k\in N_j} (\mathbf{n}_k\cdot \mathbf{v}_j + d_j)^2,
\end{equation}
where $j$ is the indexing of the patches and $\mathbf{v}_j$, $d_j$ are unit normal and distance from the origin of target plane $H_j$ for patch $U_j$. To avoid trivial solutions, we introduce the following unit length constraint on the plane normals $\mathbf{v}_j$ in the form of an additional energy term,
\begin{equation*}
\sum\limits_{j} (\vw_j^2 - 1)^2.
\end{equation*}
Additionally, the surface normals $\nw_k$ are computed as
\begin{equation*}
\nw_k^{(m)} = \frac{\Sw_u^{(m)}\times \Sw_v^{(m)}}{\|\Sw_u^{(m-1)}\times \Sw_v^{(m-1)}\|},
\end{equation*}
where $a^{(m)}$ denotes the value of variable $a$ at iteration step $m$ in our iterative optimization process. We use the constant norm $\|\Sw_u^{(m-1)}\times \Sw_v^{(m-1)}\|$ from the previous iteration when normalizing the current vector $\Sw_u^{(m)}\times \Sw_v^{(m)}$ for the computation of the surface normal $\nw_k$. This is standard practice to ensure that the objective function is polynomial.

All the above lead to an energy term of the form
\begin{equation}\label{eq: developability energy term}
E_{\text{d}} = \sum\limits_j\sum\limits_{\mathbf{n}_k\in N_j} (\mathbf{n}_k\cdot \mathbf{v}_j + d_j)^2 + \lambda_1 \sum\limits_j ( \vw_j^2 - 1 )^2,
\end{equation}
where $\lambda_1$ is an appropriate weight for the unit length constraint.

The importance of having patches that are overlapping, or equivalently neighboring patches containing common sample points, becomes evident at this point. Each patch is optimized to have a Gauss image which is a subset of a spherical curve. This can have a competitive effect between patches that are adjacent due to diverging target planes, and cause slow convergence. By having the patches share sample points, we introduce a diffusion factor to the optimization that ensures smoothness of the resulting Gauss image curve.
\\

\noindent\textbf{Soft constraints.} We also introduce a set of additional energy terms to the main problem that constrain the output surface and aim to avoid degeneracies, produce more aesthetically pleasing results and give control to the user over the proximity of the resulting surface to a reference surface.

The energy term $E_{\text{c}}$ denotes a measure of the closeness of the resulting surface $\mathbf{S}$ to a reference surface $\mathbf{S}_{\text{ref}}$, which can be either an arbitrary surface or the initial configuration of the design surface. The implementation we follow for the closeness energy term is based on the \emph{tangential distance minimization} (TDM) \cite{Pottmann:2004:RWI:1024498.1024501,Wang:2006:FBC:1138450.1138453}. The energy term is defined as the sum 
of squared distances of sample points to the tangent planes at their closest points on the reference surface. 
We use the already sampled points $\mathbf{p}_k \in \mathbf{S}$ and a set of sample points $X$ from the reference surface $\mathbf{S}_{\text{ref}}$. If the reference surface is the initial surface then $X=\{\mathbf{p}_k\}$; otherwise, $X$ is an independent sampling. Then $E_{\text{c}}$ is defined as
\begin{equation}\label{eq: closeness energy term}
E_{\text{c}} =  \sum\limits_{k} [(\mathbf{p}_k − \mathbf{x}_k)\cdot \mathbf{N}(\mathbf{x}_k)]^2,
\end{equation}
where $\mathbf{x}_k$ is the closest point to $\mathbf{p}_k$ from the set of points $X$ in the Euclidean metric, and $\mathbf{N}(\mathbf{x}_k)$ is the unit normal of $\mathbf{S}_{\text{ref}}$ at point $\mathbf{x}_k$. At each iteration the closest point is updated. We utilize FLANN for the closest point query and refer to  \cite{flann_pami_2014} for the computational complexity.

A final fairness energy term $E_{\text{f}} = w_{\text{f}_1}E_{\text{f}_1} + w_{\text{f}_2}E_{\text{f}_2}$ is introduced to the objective function that avoids degeneracies in the resulting surface  and is widely used in mesh optimization problems for the smoothing effect it provides. Specifically, we denote by $E_{\text{f}_1}$ the sum of squared norms of the first order differences of the control points in both grid directions, and by $E_{\text{f}_2}$ the second order equivalent, namely

\begin{equation*}
\begin{aligned}
& E_{\text{f}_1}=&& \sum\limits_{i,j}\left(\| \mathbf{P}_{i+1,j} - \mathbf{P}_{i,j} \|^2 + \| \mathbf{P}_{i,j+1} - \mathbf{P}_{i,j} \|^2\right), \\
& E_{\text{f}_2} =&& \sum\limits_{i,j}\left(\| \mathbf{P}_{i+1,j} - 2\mathbf{P}_{i,j} + \mathbf{P}_{i-1,j} \|^2 + \| \mathbf{P}_{i,j+1} - 2\mathbf{P}_{i,j} + \mathbf{P}_{i,j-1} \|^2\right).
\end{aligned}
\end{equation*}
We assign $w_{\text{f}_1}=0$, $w_{\text{f}_2}=1$ in all the following applications unless stated otherwise.\\

\noindent\textbf{Total energy.} All energy terms $E_{\text{d}}$, $E_{\text{c}}$, $E_{\text{f}}$ are
assigned weights $w_\text{d}$, $w_\text{c}$, $w_\text{f}$ and collected in the total energy for
developability optimization,
\begin{equation}\label{eq: total energy}
E = w_\text{d}E_{\text{d}} + w_\text{c}E_{\text{c}} + w_\text{f}E_{\text{f}}.
\end{equation}
For details on the choice of weights, we refer to Section~\ref{sec: experiments}.\\

\noindent\textbf{Increasing developability.} Now our problem is reduced to the minimization of $E$. 

\mathleft
\begin{algorithm}[H]
\caption{Increasing developability}
\label{op: main optimization problem}
\begin{equation*}
\text{minimize} \quad E = w_\text{d}E_{\text{d}} + w_\text{c}E_{\text{c}} + w_\text{f}E_{\text{f}}
\end{equation*}
\end{algorithm}
\mathcenter

The variables of $E$ are the control points $\{\mathbf{P}_{i,j}\}$ of $\mathbf{S}$ and the patch planes $H_j$, defined by $\mathbf{v}_j$ and $d_j$. The optimization problem \ref{op: main optimization problem} is an unconstrained nonlinear least-squares problem. Any algorithm for nonlinear least-squares problem can be applied in our case. We follow the standard \emph{Gauss-Newton method} in our implementation and experiments \cite[Section 10.3]{nocedal2006numerical}.


\section{Panelization}\label{sec: panelization}

\begin{figure}[H]
	\floatbox[{\capbeside\thisfloatsetup{capbesideposition={right,bottom},capbesidewidth=0.45\linewidth}}]{figure}[\linewidth]
	{\caption{Side detail of \emph{Nur Alem}, the main pavilion of the Astana EXPO 2017 Exhibition in Astana, Kazakhstan. Mostly cylindrical panels were used to rationalize the curved transparent freeform façade
			(different from the sphere).\\~\\~\\~\\}\label{fig: astana}}
	{\hspace{-27.2pt}
		\begin{tikzpicture}
		\node[anchor=south west,inner sep=0] (image) at (0,0) {\hspace{8pt}\includegraphics[height=125pt, width=\linewidth]{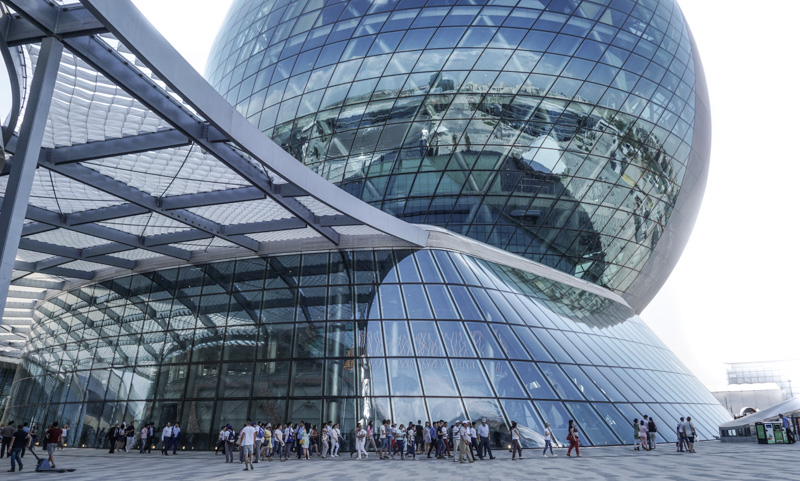}};
		\begin{scope}[x={(image.south east)},y={(image.north west)}]
		\end{scope}
		\end{tikzpicture}}
\end{figure}

Motivated by applications in architecture, we consider the problem of approximating a given arbitrary surface by a $C^0$ continuous surface which consists of developable patches. 
As we optimize for developability with help of a planar Gauss image, the resulting surface patches
include as important special cases rotational cylinders and rotational cones. We will particularly
focus on the constraints which ensure that we obtain these special types of panels. Especially
when working with glass, these rotational panels are preferred because there are special
machines for their production. Figure \ref{fig: astana} shows a recent example of an architectural freeform
facade which has been constructed with mainly cylindrical glass panels to reduce manufacturing cost.


\vspace{-148.8pt}
\noindent\hrulefill
\vspace{113.3pt}
\par\noindent\hrulefill\\


In this section, we will go through the differences between the central method that was presented in the previous sections and the variation for this new problem while introducing any new concepts that will be of use.\\

\noindent\textbf{Surface.} The main object of study in this section will be a surface $\mathbf{S}$ consisting of a grid of subsurfaces $\mathbf{S}^{(r)}$, with $C^0$ continuity at the inner boundaries. Specifically, $\mathbf{S}$ is a composite surface
$$
\mathbf{S}=\underset{r}\cup\mathbf{S}^{(r)},
$$
where $r$ indexes the set of subsurfaces. Each $\mathbf{S}^{(r)}$ is a bicubic B\'ezier surface of the form
\begin{equation}\label{eq: panel surface}
\mathbf{S}^{(r)}(u,v)=\sum\limits_{i=0}^{3}\sum\limits_{j=0}^3 B_{i,3}(u) B_{j,3}(v) \mathbf{P}_{i,j}^{(r)}
\end{equation}
and will be referred to as a \emph{panel} in the following. This configuration represents the paneling of a freeform surface. The $C^0$ continuity of neighboring panels is achieved by having common control points at the corresponding edges and models the connectivity and continuity found between distinct panels of a panelized surface. This allows for panelizations that are not in a grid configuration and can easily generalize to more complex surfaces of arbitrary topology just by appropriately "gluing" panels at their edges.\\

\noindent\textbf{Sampling and grouping.} We sample surface $\mathbf{S}$ at a collection of sample points $\{\mathbf{p}_k\}$ and group them to groups $U_r$, each corresponding to a single panel. The same follows for the corresponding surface normals $N_r$ and the associated panel planes $H_r$.

We can therefore define the developability energy term per panel as 
\begin{equation}\label{eq: panel developability energy term}
E_{\text{d}}^{(r)} = \sum\limits_{\mathbf{n}_k\in N_r} (\mathbf{n}_k\cdot \mathbf{v}_r + d_r)^2 + \lambda_1 ( \vw_r^2 - 1 )^2,
\end{equation}
where $\lambda_1$ is an appropriate weight, and the developability energy term of surface $\mathbf{S}$ as
\begin{equation}
E_{\text{d}} = \sum\limits_r E_{\text{d}}^{(r)}.
\end{equation}

This modified grouping of the sample points $\{\mathbf{p}_k\}$ allows for the individual optimization of each panel, which will be studied in more detail in a following section. \\

\noindent\textbf{Rotational panels.} By introducing the additional constraint that the panel should be a rotational surface, we are optimizing for the panels to be either rotational cones or rotational cylinders. Rotational surfaces have the property that the surface normal lines are coplanar with the axis of rotation. 
Let $L_1$, $L_2$ be two lines in $\mathbb{R}^3$ with Pl\"ucker coordinates $(\mathbf{a}, \mathbf{\bar{a}})$, $(\mathbf{b}, \mathbf{\bar{b}})\in\mathbb{R}^6$ respectively. The two lines are coplanar if their Pl\"ucker
coordinates satisfy the condition 
\begin{equation}
\mathbf{a}\cdot\mathbf{\bar{b}} + \mathbf{\bar{a}}\cdot\mathbf{b} = 0. 
\end{equation}
Recall that the Pl\"ucker coordinates $(\mathbf{a}, \mathbf{\bar{a}})\in\mathbb{R}^6$ 
of a line $L\subset\mathbb{R}^3$ are given by the direction vector $\mathbf{a}\in\mathbb{R}^3$ and the moment vector $\mathbf{\bar{a}}=\mathbf{p}\times \mathbf{a}\in\mathbb{R}^3$, where $\mathbf{p}\in\mathbb{R}^3$ is a point on $L$. Obviously, these coordinates are not independent, but satisfy the Pl\"ucker
condition $\mathbf{a} \cdot \mathbf{\bar{a}}=0.$ For more information on line geometry and relevant applications, we refer to the literature \cite[Section 2.1]{pottwall:2001}.

Consider now the Pl\"ucker coordinates $(\mathbf{n}_k, \mathbf{\bar{n}}_k)\in\mathbb{R}^6$  of the  normal lines at the sample points of a panel $U_r$ and of the unknown axis of rotation $(\mathbf{a}_r, \mathbf{\bar{a}}_r)\in\mathbb{R}^6$. The desired property that the panel is a rotational surface can be expressed as  $\mathbf{a}_r\cdot\mathbf{\bar{n}}_k + \mathbf{n}_k\cdot\mathbf{\bar{a}}_r=0$ $\forall \mathbf{n}_k\in N_r$. Thus, the problem of optimizing for rotational surface panels can be formulated as minimizing the energy
\begin{equation}\label{eq: rotationality energy term before}
\sum\limits_r \sum\limits_{\mathbf{n}_k\in N_r} (\mathbf{a}_r\cdot\mathbf{\bar{n}}_k + \mathbf{n}_k\cdot\mathbf{\bar{a}}_r)^2,
\end{equation}
under the constraint that $(\mathbf{a}_r, \mathbf{\bar{a}}_r)$ describe a line, i.e., satisfy the 
Pl\"ucker condition 
$ \mathbf{a}_r \cdot \mathbf{\bar{a}}_r =0,$
and the unit length constraint $ \mathbf{a}_r^2 = 1$  on the axis direction $\aw_r$.

At this point, we focus on the fact that for a rotational panel $S^{(r)}$ with planar Gauss image, the normal $\vw_r$ of the plane $H_r$ containing the Gauss image and the direction of the rotation axis $\aw_r$ coincide. Using this fact, we denote the 
Pl\"ucker coordinates of the rotation axis by $(\vw_r, \mathbf{\bar{v}}_r)$.

By making this adaptation, we have covered the unit length constraint on the rotation axis direction by the corresponding constraint on the target plane normal in \eqref{eq: panel developability energy term}. The Pl\"ucker condition is added as an additional energy term with an appropriate weight $\lambda_2$. Considering all the above, the resulting rotationality energy term $E_{\text{r}}$ is of the form
\begin{equation}\label{eq: rotationality energy term}
E_{\text{r}} = \sum\limits_r \sum\limits_{\mathbf{n}_k\in N_r} (\mathbf{v}_r\cdot\mathbf{\bar{n}}_k + \mathbf{n}_k\cdot\mathbf{\bar{v}}_r)^2 + \lambda_2\sum\limits_r (\vw_r\cdot \mathbf{\bar{v}}_r)^2.
\end{equation}

While the Pl\"ucker coordinates of the normal lines are initialized in the optimization problem with their current values in the configuration of surface $\mathbf{S}$, the axis of rotation $(\mathbf{v}_r, \mathbf{\bar{v}}_r)$ of every panel $U_r$ remains unknown at this point or, assuming the panels are in generic configuration, does not exist at all. An appropriate initialization for the 
Pl\"ucker coordinates of the axis of rotation of each panel is given by methods used in kinematic surface reconstruction applications, where the problem of fitting a velocity field to a set of surface normals is studied 
\cite{Liu:2006:CSR:1649591.1649907,Pottmann1998}. It follows the same thought process as the main idea behind the energy term \eqref{eq: rotationality energy term before}. In fact, it is exactly the same energy that we aim to minimize but applied to each of the panels separately while considering the affine normal lines fixed. The resulting axis is the best fitting one in the least-squares sense. Formulating the above as an optimization problem leads us to the minimization of 
\begin{equation}
\sum\limits_{\mathbf{n}_k\in N_r} (\mathbf{v}_r\cdot\mathbf{\bar{n}}_k + \mathbf{\bar{v}}_r\cdot\mathbf{n}_k)^2. \label{axis-init}
\end{equation}
We already have an appropriate initialization for the target plane normal $\vw_r$, described in optimization problem \ref{op: plane initialization optimization problem}. Thus, the objective function (\ref{axis-init}) is a quadratic function of the moment vector $\mathbf{\bar{v}}_r$. The latter is orthogonal to $\vw_r$ and therefore can be expressed as 
$$\mathbf{\bar{v}}_r = \mu_1 \bw_1 + \mu_2 \bw_2, $$
where $\bw_1,\bw_2\in \mathbb{R}^3$ form a basis of the plane perpendicular to $\vw_r$. Substitution into (\ref{axis-init})
yields a quadratic function in $\mu_1,\mu_2$ and the optimal values of $\mu_1,\mu_2$ are the solutions of a linear system.\\


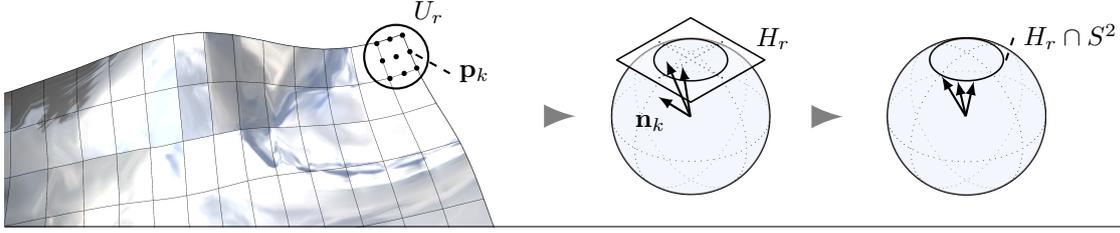
\begin{figure}[t]
\begin{subfigure}[b]{0.5\textwidth}
       \begin{tikzpicture}
          \node[anchor=south west,inner sep=0] (image) at (0,0) {\includegraphics[width=\linewidth]{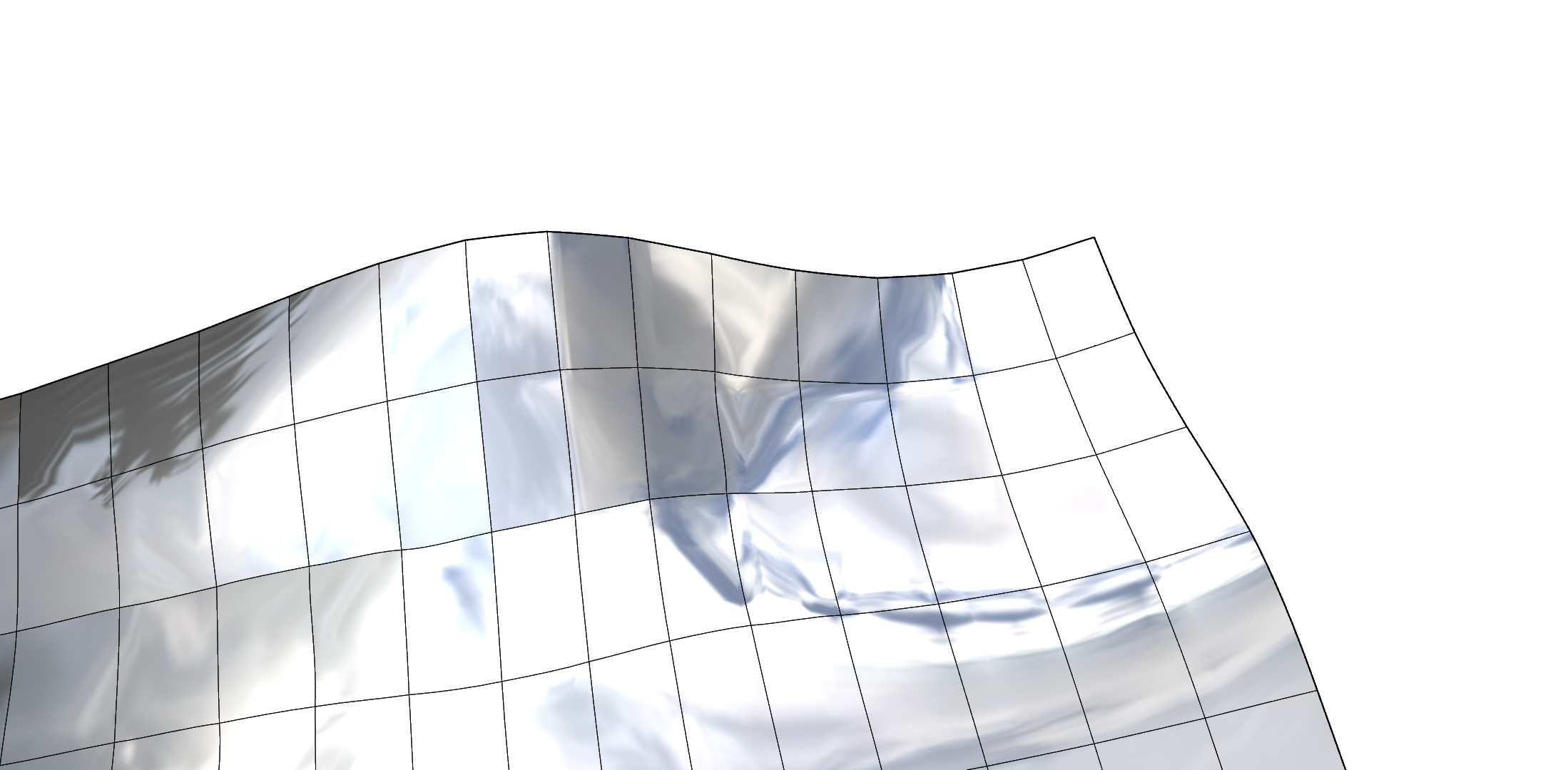}};
          \begin{scope}[x={(image.south east)},y={(image.north west)}]

          \foreach \Point in {(0.697,0.69), (0.71,0.631), (0.722,0.572), (0.7,0.553), (0.675,0.535), (0.662,0.598), (0.652,0.66), (0.675,0.676), (0.686,0.613)}{
              \fill \Point circle[radius=1pt];
          }

          \draw[thick] (0.686,0.613) circle[ radius=12pt];

          \node[above right] at (0.7,0.7) {$U_r$};

          \draw[dashed,-,thick](0.72,0.621) -- (0.78,0.553);
          \node[right] at (0.78,0.553) {$\mathbf{p}_k$};
          
          \draw[->, >=latex, line width=3pt, color=gray] (0.95,0.4) -- (1,0.4);
          

          \end{scope}
      \end{tikzpicture}
    \end{subfigure}\hspace{10pt}
    \begin{subfigure}[b]{0.14\textwidth}
    	\begin{tikzpicture}
            \node[anchor=south west,inner sep=0] (image) at (0,0) {\includegraphics[width=\linewidth]{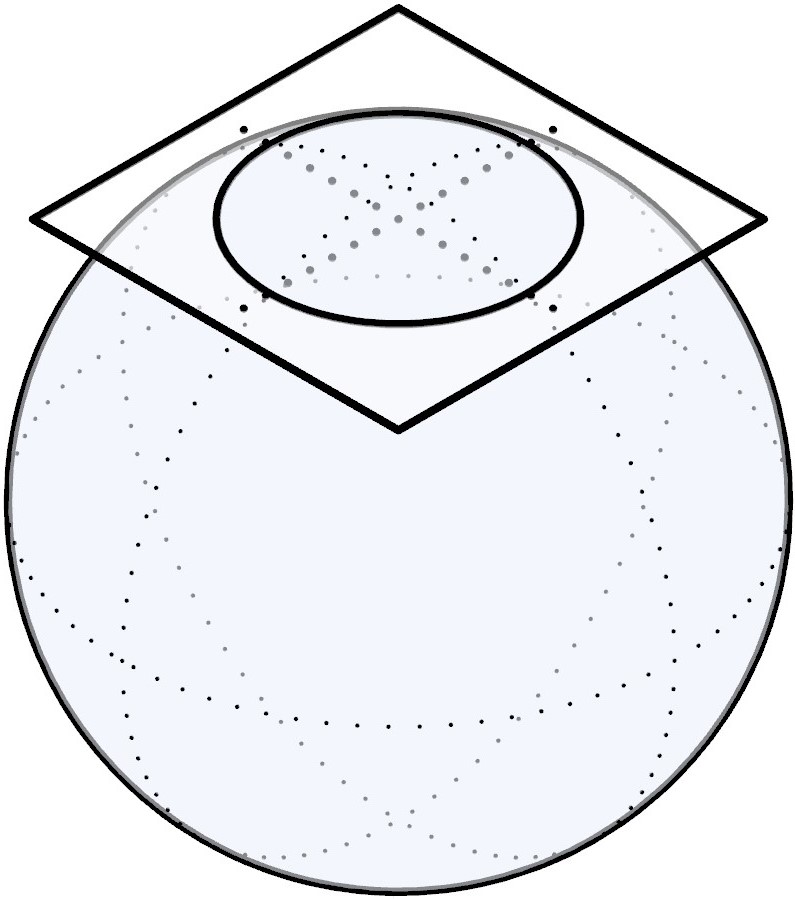}};
            \begin{scope}[x={(image.south east)},y={(image.north west)}]
            
            \draw[->,>=latex,thick](0.5,0.44) -- (0.3,0.55);
            \draw[->,>=latex,thick](0.5,0.44) -- (0.35,0.75);
            \draw[->,>=latex,thick](0.5,0.44) -- (0.45,0.72);
            
            \node[below] at (0.25,0.5) {$\mathbf{n}_k$};

            \node[above right] at (0.85,0.76) {$H_r$};
            
            \draw[->, >=latex, line width=3pt, color=gray] (1.36,0.44) -- (1.45,0.44);

            \end{scope}
        \end{tikzpicture}
        \vspace{0pt}
    \end{subfigure}\hspace{40pt}
     \begin{subfigure}[b]{0.14\textwidth}
    	\begin{tikzpicture}
            \node[anchor=south west,inner sep=0] (image) at (0,0) {\includegraphics[width=\linewidth]{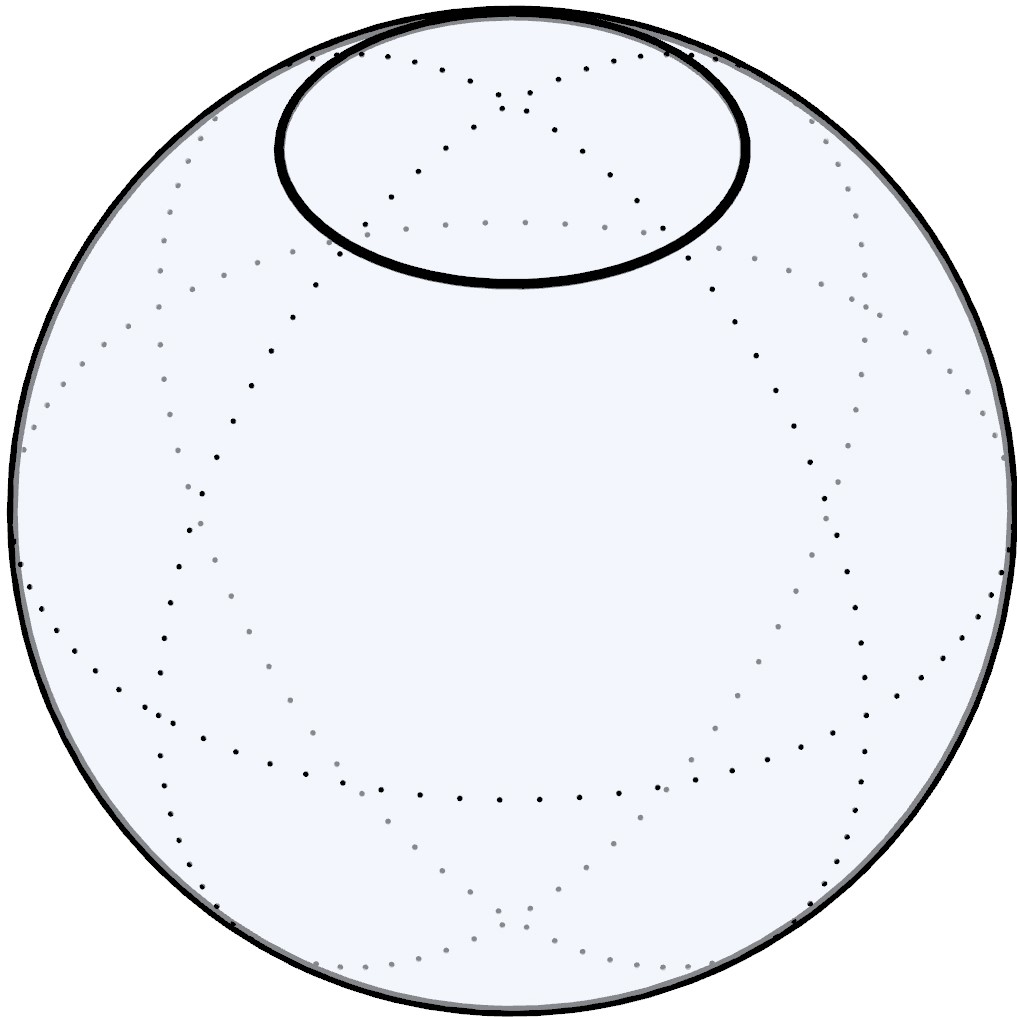}};
            \begin{scope}[x={(image.south east)},y={(image.north west)}]
            
            \draw[->,>=latex,thick](0.5,0.5) -- (0.45,0.72);
            \draw[->,>=latex,thick](0.5,0.5) -- (0.35,0.751);
            \draw[->,>=latex,thick](0.5,0.5) -- (0.55,0.72);
            
            \draw[dashed, -,thick](0.75,0.85) -- (0.8,1);
            \node[right] at (0.8,1) {$H_r\cap S^2$};

            \end{scope}
        \end{tikzpicture}
        \vspace{0pt}
    \end{subfigure}\hfill
\vspace{-12pt}
\par\noindent\hrulefill\\
\caption{We focus on a single panel $\mathbf{S}^{(r)}$ of a panelized surface $\mathbf{S}$. We are optimizing for the endpoints of normals $\mathbf{n}_k$ corresponding to the sample points $\mathbf{p}_k\in U_r$ of panel $\mathbf{S}^{(r)}$ to lie on the same plane $H_r$.}
\label{fig: single panel}
\end{figure}


\noindent\textbf{Surface paneling.} Thus, the surface paneling problem is the following variation of the optimization problem \ref{op: main optimization problem}, and is solved with the same approach.

\mathleft
\begin{algorithm}[H]
\caption{Surface paneling}
\label{op: surface paneling optimization problem}
\begin{equation*}
\text{minimize} \quad E = w_\text{d}E_{\text{d}} +w_\text{r}E_{\text{r}} + w_\text{c}E_{\text{c}} + w_\text{f}E_{\text{f}}
\end{equation*}
\end{algorithm}
\mathcenter

\noindent\textbf{Individual panel treatment.} 
Until now we have shown how to optimize the paneling of surface $\mathbf{S}$ in a global fashion. Since we defined the energy term $E_{\text{d}}^{(r)}$ per panel, this approach can be customized to consider each panel separately, achieving in the process increased control over the resulting panelization.
We use the following obvious fact:


\begin{lemma}\label{le: panel types}
Let panel $\mathbf{S}^{(r)}$ be a rotational surface and $H_r$ be a plane such that the Gauss image of the panel is entirely contained in plane $H_r$. Then the panel type is determined by the distance $d_r$ of plane $H_r$ from the origin $O$. Specifically,
\begin{enumerate}
\item If $d_r=1$ then $\mathbf{S}^{(r)}$ is planar.
\item If $d_r=0$ then $\mathbf{S}^{(r)}$ is a cylinder of revolution.
\item If $d_r\in (0,1)$ then $\mathbf{S}^{(r)}$ is a cone of revolution whose rulings form the angle $\arcsin d_r$ with
the rotation axis. 
\end{enumerate}
\end{lemma}

This offers a good way to aim at cylindrical panels or conical panels with prescribed opening angle by prescribing the according 
values of $d_r$ in the energy term $E_{\text{d}}^{(r)}$ in (\ref{eq: panel developability energy term}). 

It is often the case in industrial applications that individual adjustments need to be made to the panelization for reasons that include aesthetics and the overall cost of the project. The advantages of the individual treatment of the panels become apparent in such cases, and the aforementioned main pavilion of the Astana EXPO 2017, shown in Figure \ref{fig: astana}, serves as an example. In that project, apart from the cylindrical panels which were the main ingredient of the panelization, double curved panels were also utilized in areas that the use of cylindrical panels would negatively affect the aesthetics of the result. Thus, by integrating a singular panel management strategy to the optimization we have the ability of dealing with isolated problematic areas without sacrificing the quality of the overall panelization.


\section{Experiments, results and discussion}\label{sec: experiments}

\begin{example}\label{ex: leather}
In this example, we consider a mesh $\mathcal{M}$ which originated from scanning a thin deformed leather patch. The deformation was introduced to the material in the form of local stretches along its surface which result in areas of nonzero Gaussian curvature.\\


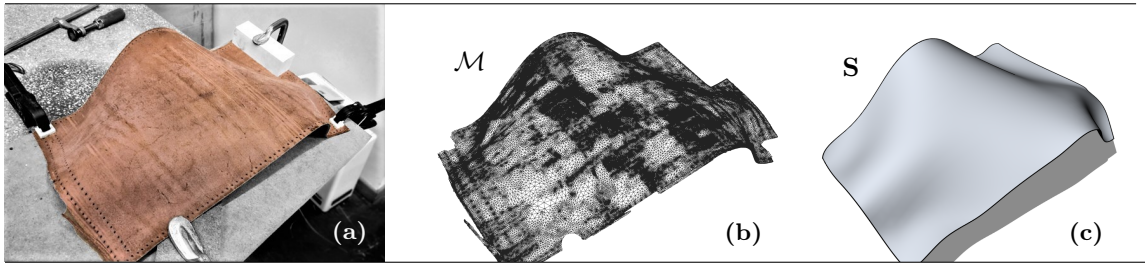
\begin{figure}[H]
\centering
\centering
\begin{subfigure}[b]{0.34\linewidth}
  \begin{tikzpicture}
    \node[anchor=south west,inner sep=0] (image) at (0,0) {\includegraphics[height=97pt]{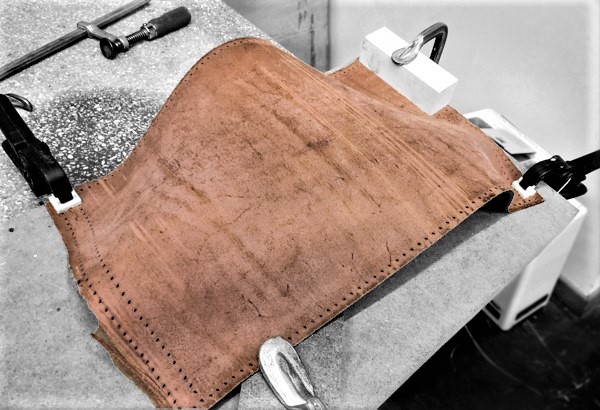}};
    \begin{scope}[x={(image.south east)},y={(image.north west)}]
    \node[above left] at (0.98,0.03) {\color{lightgray!10}\small\textbf{(a)}};
    \end{scope}
  \end{tikzpicture}%
  \label{subfig: leather}
\end{subfigure}%
\hfill 
\begin{subfigure}[b]{0.3\linewidth}
  \begin{tikzpicture}
    \node[anchor=south west,inner sep=0] (image) at (0,0) {\includegraphics[trim={0 50pt 0 0},clip,width=\linewidth]{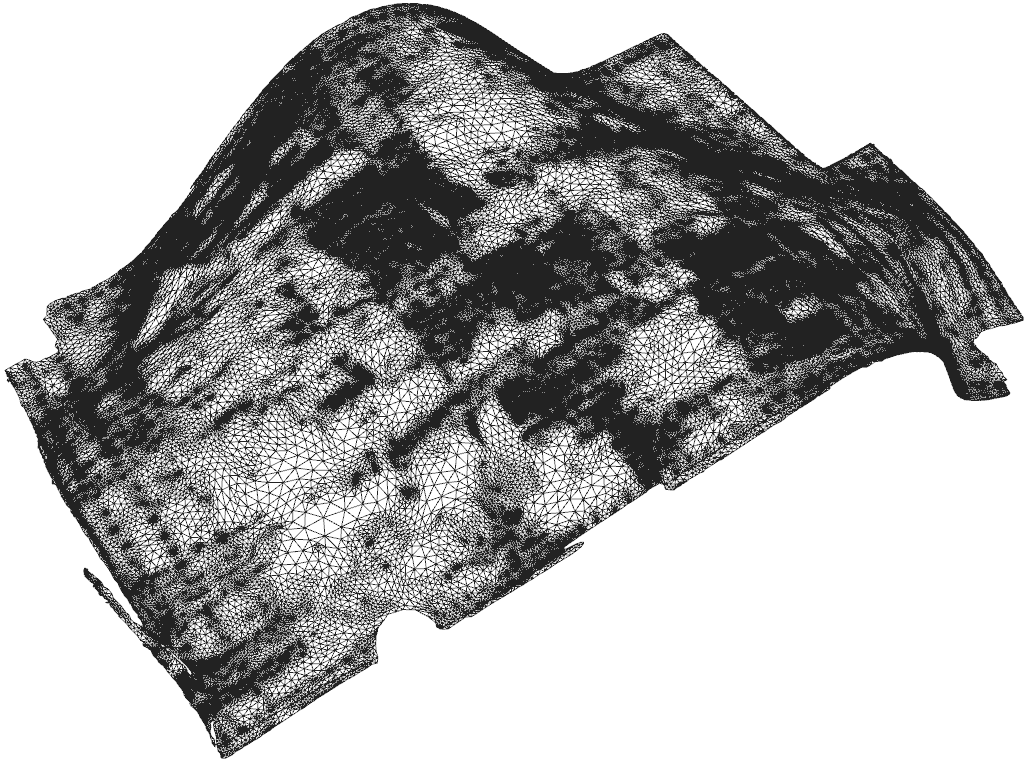}};
    \begin{scope}[x={(image.south east)},y={(image.north west)}]
    \node[below] at (0.1,0.95) {$\mathcal{M}$};
    \node[above left] at (0.98,0.03) {\small\textbf{(b)}};
    \end{scope}
  \end{tikzpicture}%
  \label{subfig: leather scan mesh}
\end{subfigure}%
\hfill 
\begin{subfigure}[b]{0.26\linewidth}
  \begin{tikzpicture}
    \node[anchor=south west,inner sep=0] (image) at (0,0) {\includegraphics[trim={0 50pt 0 0},clip,width=\linewidth]{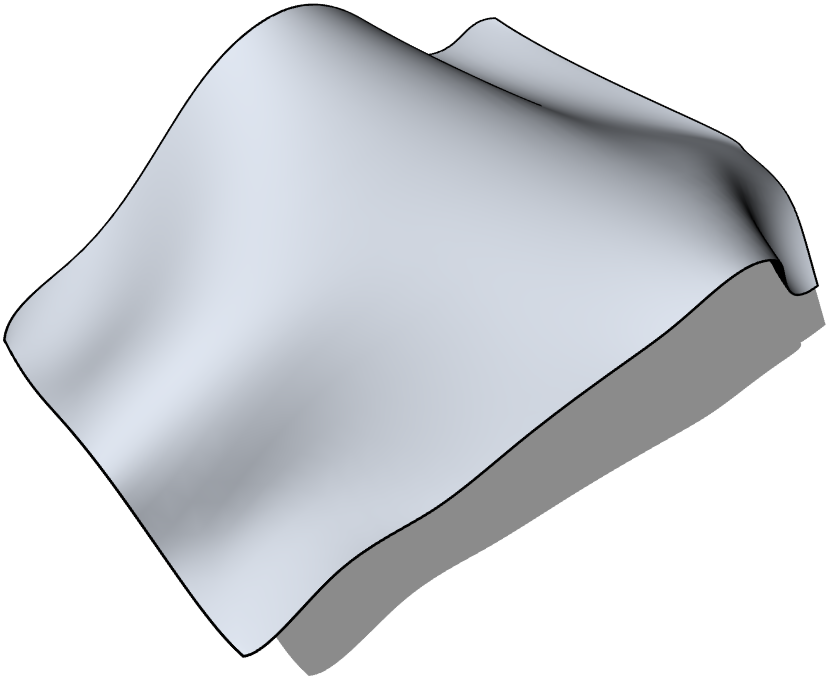}};
    \begin{scope}[x={(image.south east)},y={(image.north west)}]
	\node[below] at (0.1,0.95) {$\mathbf{S}$};
    \node[above left] at (0.98,0.03) {\small\textbf{(c)}};
    \end{scope}
  \end{tikzpicture}%
  \label{subfig: surface}
\end{subfigure}%
\quad
\vspace{-109pt}
\par\noindent\hrulefill
\vspace{97pt}
\vspace{-11.5pt}
\par\noindent\hrulefill
\caption{\textbf{(a)} The configuration of the deformed leather patch. \textbf{(b)} Mesh acquired from scanning the leather material. \textbf{(c)} The material's geometry is represented as a B-spline surface.}\label{fig:leather}
\end{figure}


To apply our algorithm for increasing developability, we first fit the data with a  bicubic B-spline surface $\mathbf{S}$ of the form (\ref{eq:surface}) with $7\times 13$ control points. 
This is done using the TDM optimization framework for surface fitting described in section \ref{subsec: problem formulation}. We refer to the initial configuration of surface $\mathbf{S}$, given by the fitting optimization, as $\mathbf{S}_0$. Following the procedure described in section \ref{sec: optimization setup}, we sample the resulting surface $\mathbf{S}$ uniformly along the parameter space at $30\times 60$ evaluation points $\mathbf{p}_{i,j},\ i\in [1,30],\ j\in [1,60]$. We then group $\mathbf{p}_{i,j}$ in patches $U_{l,m}$, each one containing $5\times 5$ points with an overlap in both directions of $2$ points between neighboring patches, i.e. $U_{l,m}=\{\mathbf{p}_{i,j}\ |\ i\in [3l-2,\ 3l+2],\ j\in [3m-2,\ 3m+2]\}$. This completes the initialization of the optimization algorithm of problem \ref{op: main optimization problem}.



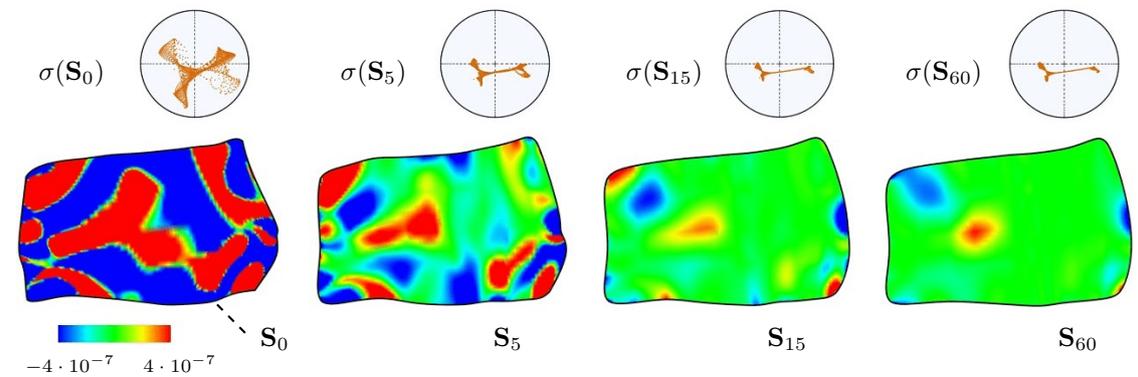
\begin{figure}[H]
\centering
  \begin{tikzpicture}
  \node[anchor=south west,inner sep=0] (image) at (0,0) {\includegraphics[width=0.98\linewidth]{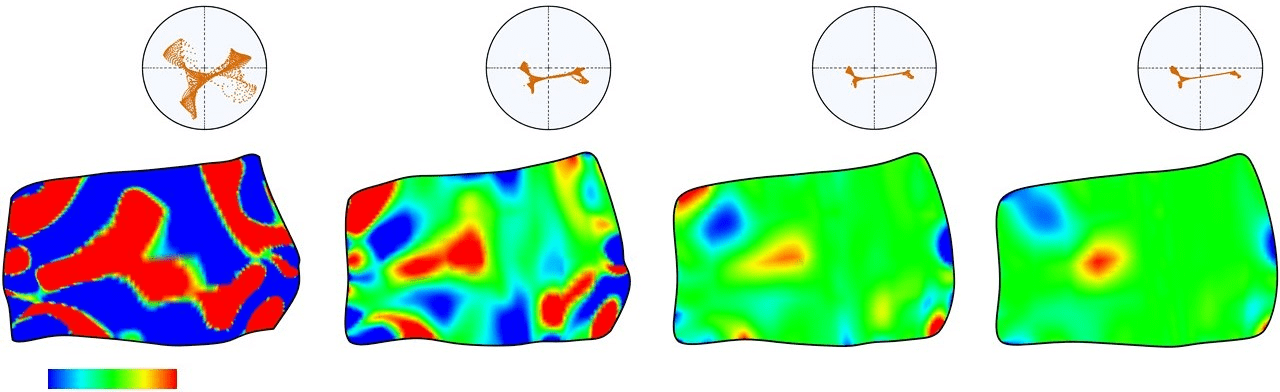}};
    \begin{scope}[x={(image.south east)},y={(image.north west)}]


      \node at (0.05,0.8) {$\sigma(\mathbf{S}_0)$};
      \node at (0.32,0.8) {$\sigma(\mathbf{S}_5)$};
      \node at (0.58,0.8) {$\sigma(\mathbf{S}_{15})$};
      \node at (0.83,0.8) {$\sigma(\mathbf{S}_{60})$};
      
      \draw[dashed,-,thick] (0.18,0.12) -- (0.21,0.02);
      \node[right] at (0.21,0.02) {$\mathbf{S}_0$};
      
      \node at (0.44,0.02) {$\mathbf{S}_5$};
      \node at (0.69,0.02) {$\mathbf{S}_{15}$};
      \node at (0.95,0.02) {$\mathbf{S}_{60}$};
      
      \node[below right] at (0,0) {\scriptsize $-4\cdot 10^{-7}$\quad \ $4\cdot 10^{-7}$};

    \end{scope}
  \end{tikzpicture}%
  \vspace{-12pt}
  \par\noindent\hrulefill
\caption{The Gauss map (top) and the Gaussian curvature (bottom) of surface $\mathbf{S}$ for different numbers of iterations, namely at 0, 5, 15 and 60 ($\mathbf{S}_t$ denotes the optimized surface at $t$ iterations). The length of the surface has been scaled to be approximately 1.}
\label{fig:iterations}
\end{figure}


We introduce to the optimization process a closeness energy term of the form (\ref{eq: closeness energy term}) with relatively small weight to ensure proximity of $\mathbf{S}$ to its original position $\mathbf{S}_{0}$. As described before, this is implemented using the TDM framework. We consider the original surface $\mathbf{S}_0$ as the reference surface and use the already sampled points $\mathbf{p}_{i,j}$ of surface $\mathbf{S}$ as the evaluation points of the TDM algorithm. In our experiments, we observed that using this competing low-weight term in our main optimization procedure constrains the solution space by avoiding trivial solutions and producing results that are more desirable from the designer's point of view.

Figure \ref{fig: surface before after} reveals the inner workings of the developability algorithm, which clearly produces a "thinner" Gauss image for the resulting surface and also illustrates a comparison between the original surface $\mathbf{S}_0$ and the resulting surface $\mathbf{S}$. Figure \ref{fig:iterations} shows the Gauss map and the Gaussian curvature of the surface for several intermediate iterations of the optimization. The detailed statistics for this example are given in Table \ref{tab:leather}.\\


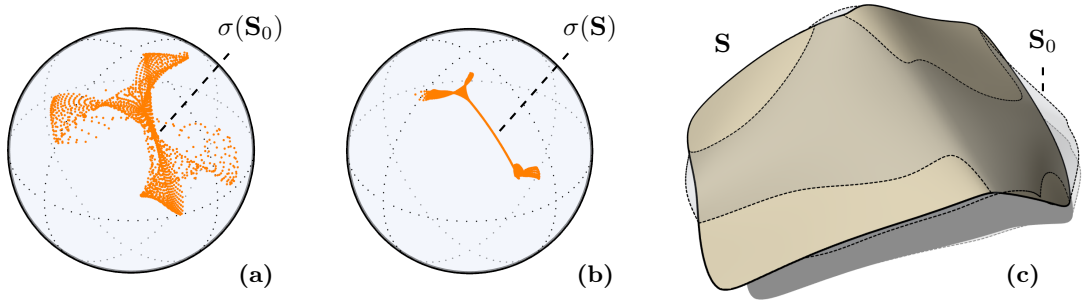
\begin{figure}[H]
\centering
\quad
\begin{subfigure}[b]{0.22\textwidth}
  \begin{tikzpicture}
    \node[anchor=south west,inner sep=0] (image) at (0,0) {\includegraphics[width=\linewidth]{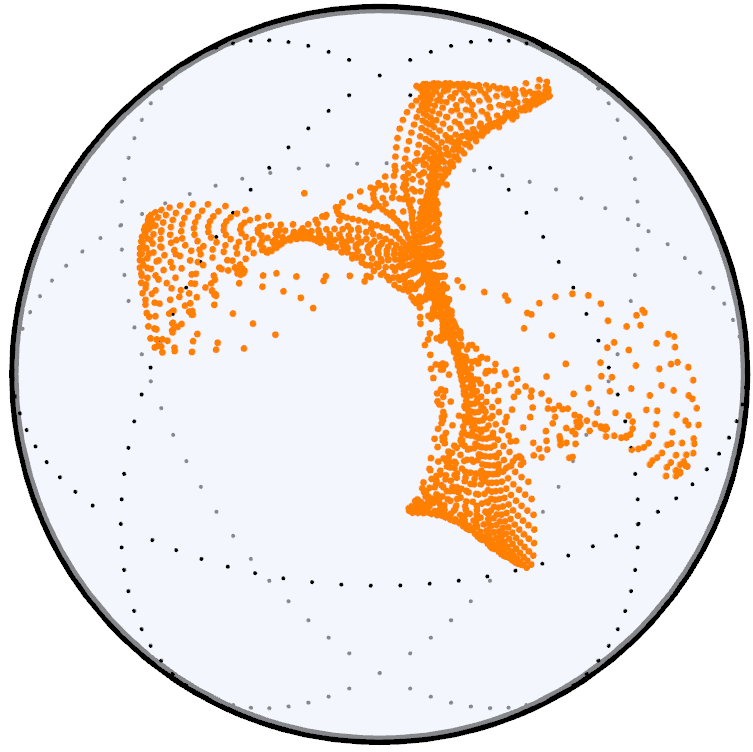}};
    \begin{scope}[x={(image.south east)},y={(image.north west)}]
    
    \node at (1,0) {\small\textbf{(a)}};
    
    \draw[dashed,-,thick] (0.62,0.58) -- (0.9,0.9);
    \node[above] at (0.98,0.9) {$\sigma(\mathbf{S}_0)$};
    
  \end{scope}
  \end{tikzpicture}%
  \vspace{0pt}
  \label{subfig: leather gauss image before}
\end{subfigure}
\hfill %
\begin{subfigure}[b]{0.22\textwidth}
  \centering
  \begin{tikzpicture}
    \node[anchor=south west,inner sep=0] (image) at (0,0) {\includegraphics[width=\linewidth]{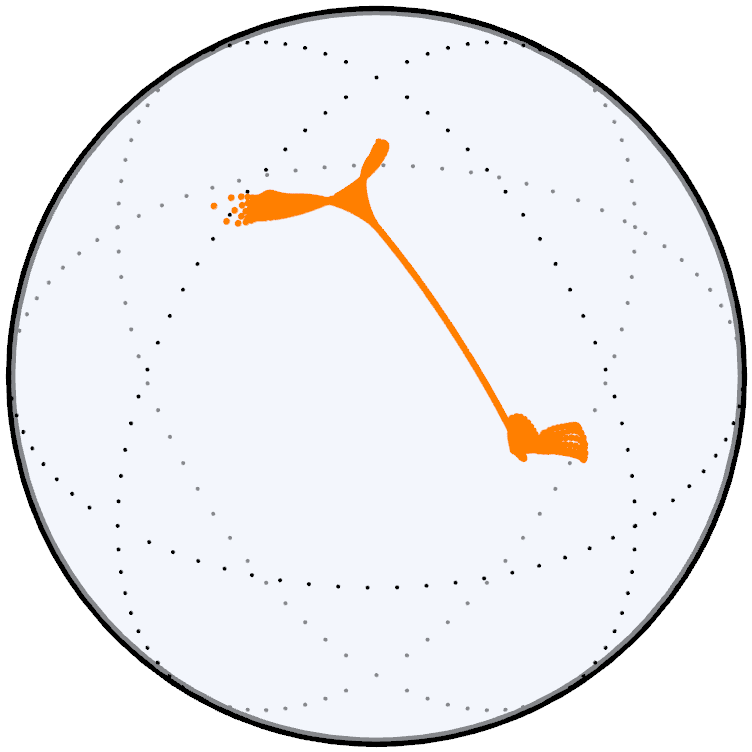}};
    \begin{scope}[x={(image.south east)},y={(image.north west)}]
    
     \draw[dashed,-,thick] (0.62,0.58) -- (0.9,0.9);
    \node[above] at (0.98,0.9) {$\sigma(\mathbf{S})$};
    
    \node at (1,0) {\small\textbf{(b)}};
    
    \end{scope}
  \end{tikzpicture}%
  \vspace{0pt}
  \label{subfig: leather gauss image after}
\end{subfigure}
\hfill %
\begin{subfigure}[b]{0.35\textwidth}
  \begin{tikzpicture}
    \node[anchor=south west,inner sep=0] (image) at (0,0) {\includegraphics[width=\textwidth]{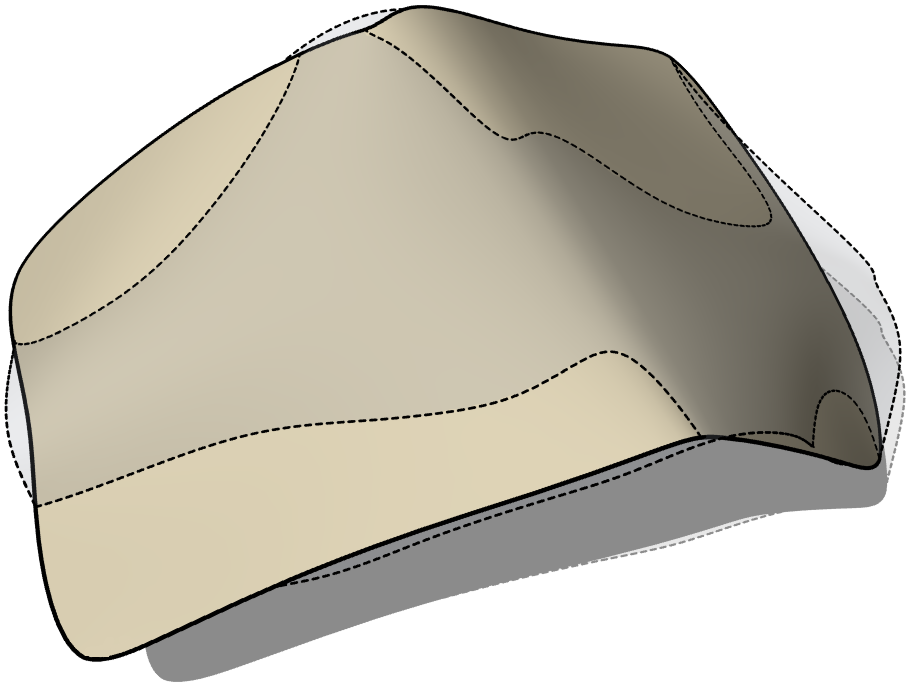}};
    \begin{scope}[x={(image.south east)},y={(image.north west)}]
    
     \draw[dashed,-,thick] (0.9,0.7) -- (0.9,0.8);
    \node[above] at (0.9,0.8) {$\mathbf{S}_0$};
    
    \node[above] at (0.1,0.8) {$\mathbf{S}$};
    
    \node at (0.85,0.08) {\small\textbf{(c)}};
    
    \end{scope}
  \end{tikzpicture}%
  \vspace{-1pt}
  \label{subfig: surface before after}
\end{subfigure}
\quad
\vspace{-11.6pt}
\par\noindent\hrulefill
\caption{\textbf{(a)} The Gauss image of the initial configuration of B-spline surface $\mathbf{S}_0$ representing the leather material. \textbf{(b)} The Gauss image of the optimized surface $\mathbf{S}$. \textbf{(c)} The optimized B-spline surface $\mathbf{S}$ in solid color compared to the transparent initial surface $\mathbf{S}_0$.}\label{fig: surface before after}
\end{figure}


We already discussed in section \ref{subsec:thin} that the straightening of one family of principal curvature lines of $\mathbf{S}$ compared to the principal curvature lines of the initial surface $\mathbf{S}_0$ is an alternative indication of the increase in developability. Figure \ref{fig:principal} demonstrates the straightening effect in this example. Also illustrated is that the preimage of a small collection of points in one of the "thinner" parts of the Gauss image corresponds to one of the approximate rulings of the surface.\\


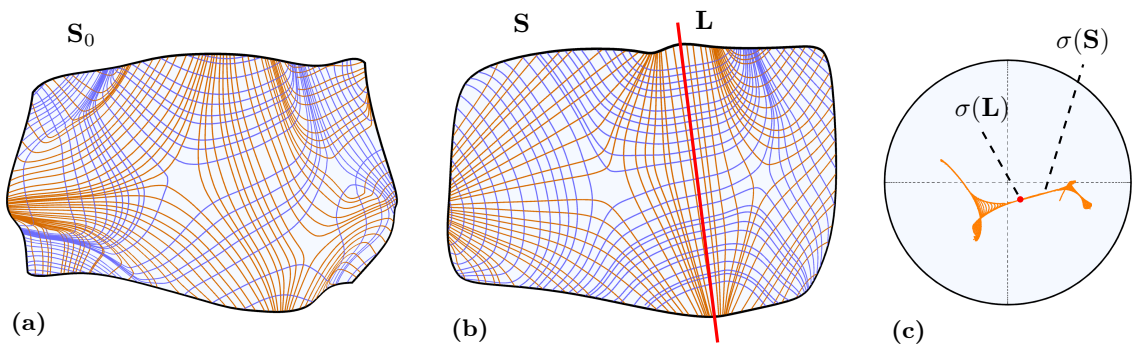
\begin{figure}[H]
\centering
\begin{subfigure}[b]{0.35\linewidth}
  \begin{tikzpicture}
    \node[anchor=south west,inner sep=0] (image) at (0,0) {\includegraphics[width=\linewidth]{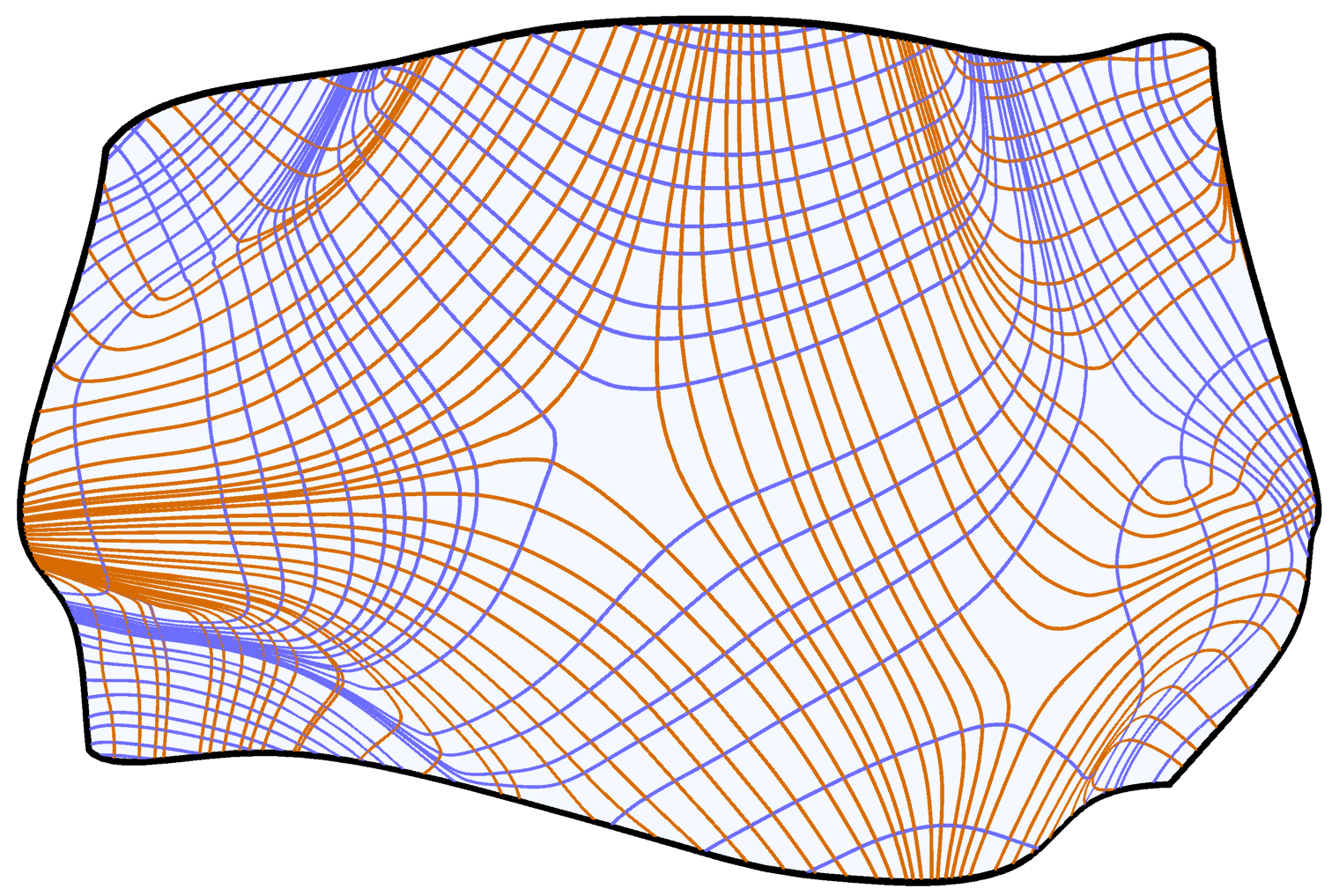}};
    \begin{scope}[x={(image.south east)},y={(image.north west)}]

    \node at (0.2,1.05) {$\mathbf{S}_0$};
    
    \node[above right] at (0,-0.11) {\small\textbf{(a)}};

    \end{scope}
  \end{tikzpicture}
  \vspace{-18.3pt}
\end{subfigure}
\hfill
\begin{subfigure}[b]{0.35\linewidth}
  \begin{tikzpicture}
    \node[anchor=south west,inner sep=0] (image) at (0,0) {\includegraphics[width=\linewidth]{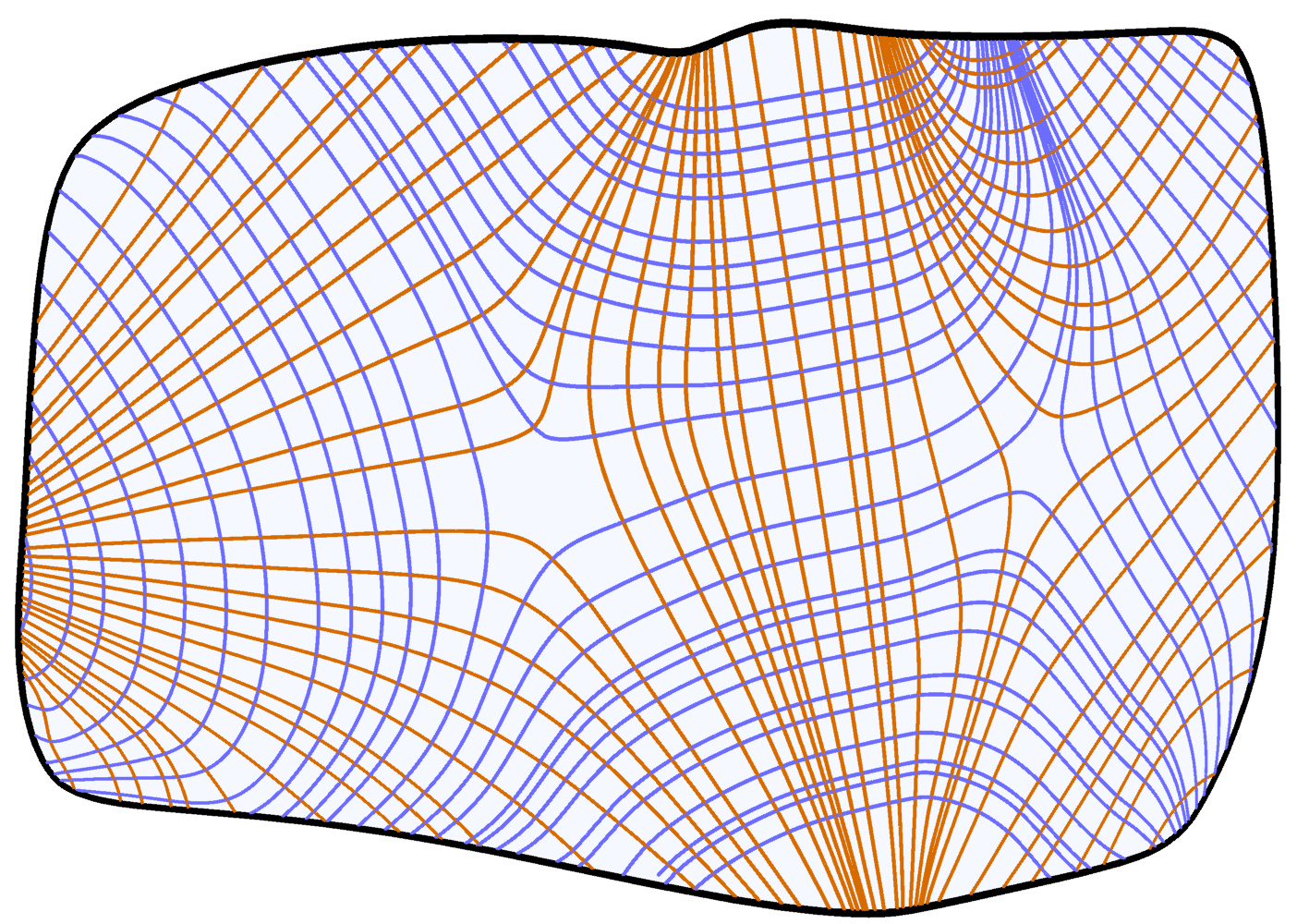}};
    \begin{scope}[x={(image.south east)},y={(image.north west)}]

    \node at (0.2,1.05) {$\mathbf{S}$};
    
    
    \draw[red,-,very thick] (0.59,1.05) -- (0.69,-0.08);
    \node[right] at (0.61,1.06) {$\mathbf{L}$};
    
    \node[above right] at (0,-0.11) {\small\textbf{(b)}};

    \end{scope}
  \end{tikzpicture}
 \vspace{-20.5pt}
\end{subfigure}
\hfill
\begin{subfigure}[b]{0.22\linewidth}
  \begin{tikzpicture}
    \node[anchor=south west,inner sep=0] (image) at (0,0) {\includegraphics[width=\linewidth]{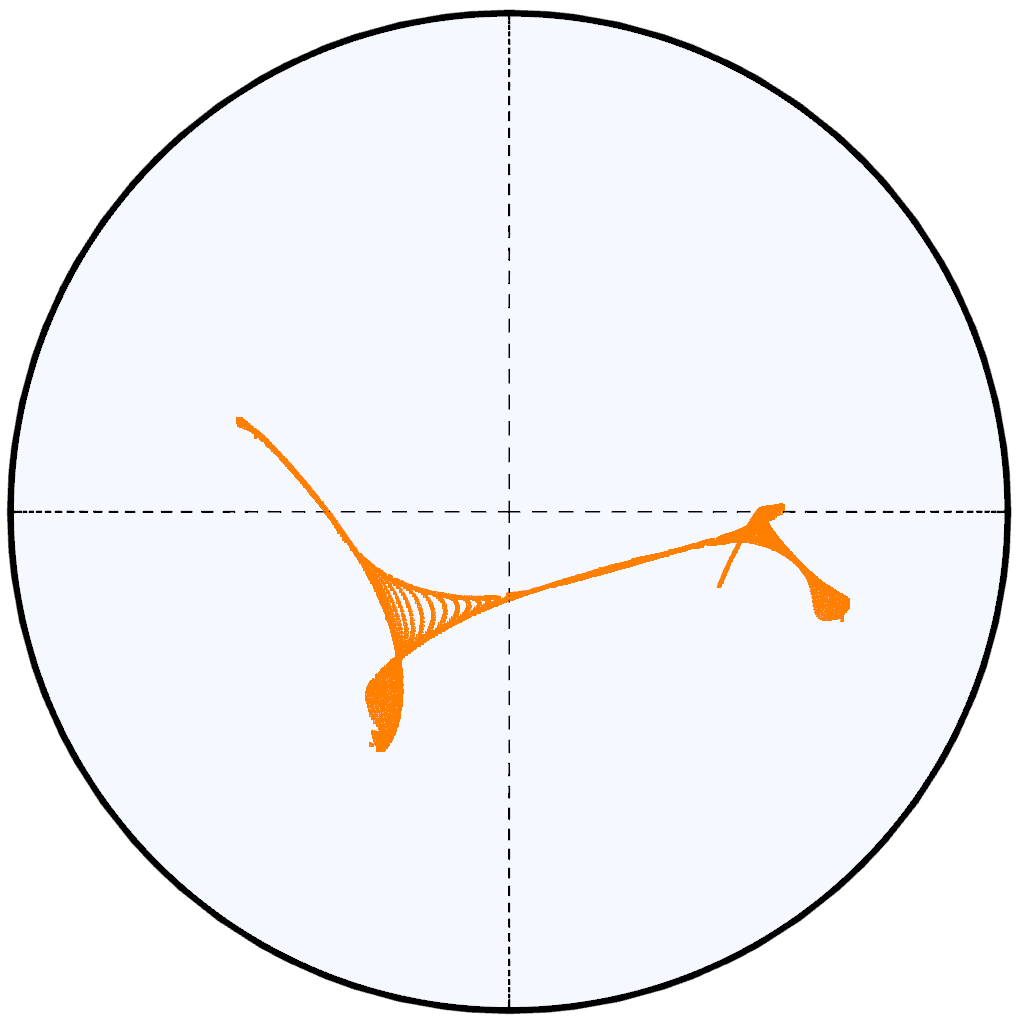}};
    \begin{scope}[x={(image.south east)},y={(image.north west)}]
    
    \draw[dashed,-,thick] (0.65,0.47) -- (0.8,0.97);
    \node[above] at (0.8,0.97) {$\sigma(\mathbf{S})$};
    
    \fill[red, thick] (0.55,0.43) circle (1.2pt);
    \draw[dashed,-,thick] (0.54,0.45) -- (0.4,0.7);
    \node[above] at (0.4,0.7) {$\sigma(\mathbf{L})$};
    
    \node[above right] at (0,-0.187) {\small\textbf{(c)}};

    \end{scope}
  \end{tikzpicture}
 \vspace{-21.5pt}
\end{subfigure}
\vspace{10pt}
\par\noindent\hrulefill
\caption{Visualization of the principal curvature lines. \textbf{(a)} The principal curvature lines of the initial surface $\mathbf{S}_0$. \textbf{(b)} The principal curvature lines of the optimized surface $\mathbf{S}$. Highlighted in red and extended slightly for clarity, one such principal curvature line $\mathbf{L}$, which also approximately corresponds to the preimage of a small collection of points around the "thin" part of $\sigma(\mathbf{S})$ . \textbf{(c)} The Gauss image $\sigma(\mathbf{S})$ of the optimized surface $\mathbf{S}$. The Gauss image of $\mathbf{L}$ is highlighted in red. }\label{fig:principal}
\end{figure}




\begin{figure}[H]
  \begin{tikzpicture}
  \node[anchor=south west,inner sep=0] (image) at (0,0) {\includegraphics[width=\linewidth]{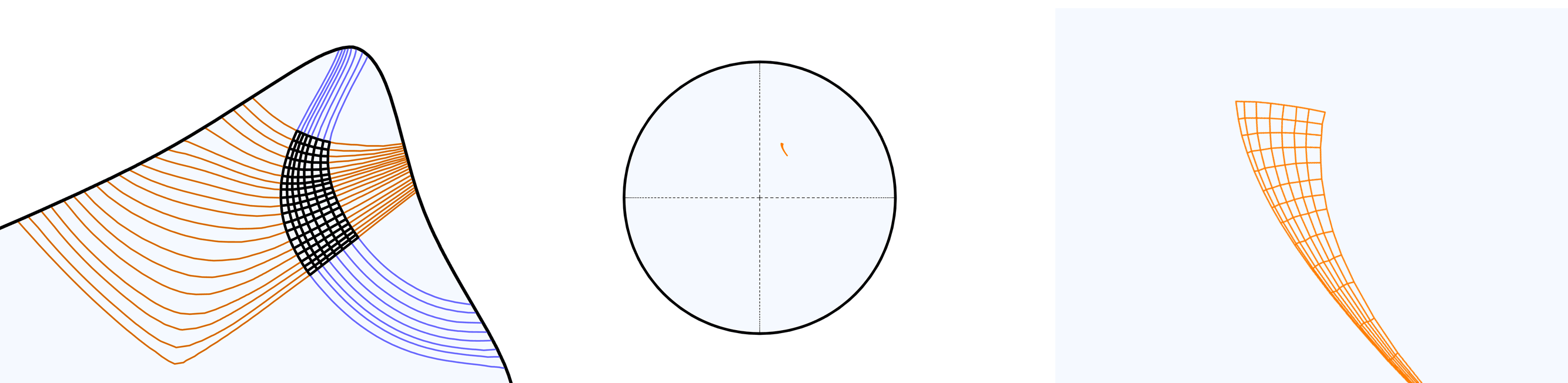}};
    \begin{scope}[x={(image.south east)},y={(image.north west)}]

      \node at (0.05,0.7) {$\Sw$};
      \node[left] at (0.42,0.8) {$S^2$};
      
      \draw[dashed,-,thick] (0.235,0.38) -- (0.31,0.38);
      \node[right] at (0.31,0.38) {$\mathcal{P}$};
      
      \node at (0.93,0.7) {$\sigma(\mathcal{P})$};
      
      \draw[densely dotted,-,thick] (0.488,0.579) rectangle (0.512,0.641);
      \draw[dotted,-] (0.488,0.641) -- (0.675,0.978);
      \draw[dotted,-] (0.488,0.579) -- (0.675,0);
      \draw[dotted,-] (0.512,0.641) -- (0.675,0.75);
      \draw[dotted,-] (0.512,0.579) -- (0.675,0.35);
      \draw[dashed,-,line width=1.2pt] (0.675,0) -- (0.675,0.978) -- (1,0.978);

    \end{scope}
  \end{tikzpicture}%
  \vspace{-12pt}
  \par\noindent\hrulefill
\caption{We consider a nearly developable patch of a surface $\Sw$ and the two families of principal curvature lines of $\Sw$ (blue and orange lines) over that patch. These families define a principal net denoted with $\mathcal{P}$. The Gauss image $\sigma(\mathcal{P})$ of the net is displayed on the right.}
\label{fig: principal net}
\end{figure}
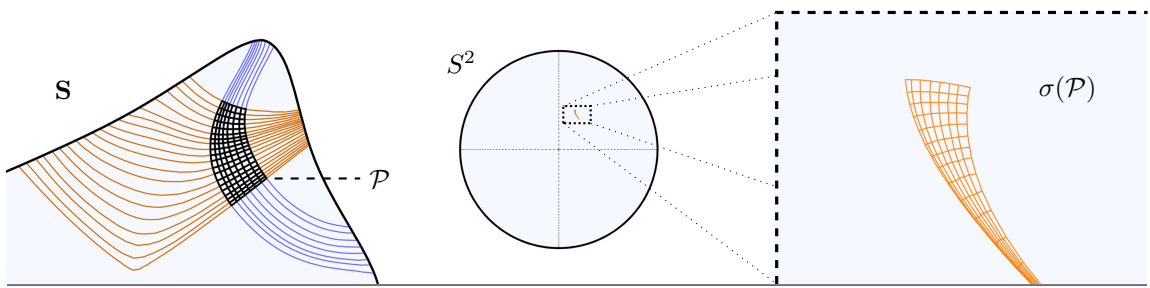



\begin{table}[H]
  \centering
  \footnotesize
  \caption{We present the detailed information for the optimization of the leather surface $\mathbf{S}$. The number of control points of $\mathbf{S}$ and the number of overlapping patches that cover the surface generate the number of variables (3 per control point and 4 per patch-associated plane). The surface was evaluated at 1800 points and each patch contained 25 points. The weights were chosen to favor the developability property. The initial and intermediate total energies of the problem were $E_0 = 9328.17$, $E_{5} = 2103.75$, $E_{15} = 356.702$ while the order of the final total energy $E_{60} = 5.08$ was achieved at iteration 26, where $E_{26} = 5.38$. Also provided, the total time, time used by the Newton solver, and the time per iteration (in seconds), measured on an Intel\textsuperscript{\tiny\textregistered} Core\textsuperscript{\tiny\texttrademark} i7-6700HQ processor.}
    \begin{tabular}{|ccc|ccc|ccc|c|ccc|}
    \hline
    \multicolumn{3}{|c|}{number of...} & \multicolumn{3}{c|}{weights} & \multicolumn{3}{c|}{final energies} & \multicolumn{1}{c|}{number of} & \multicolumn{3}{c|}{time [sec]}\\
    \multicolumn{1}{|l}{ctrl.pts} & \multicolumn{1}{l}{patches} & \multicolumn{1}{l|}{variables} & \multicolumn{1}{l}{$w_{\text{d}}$} & \multicolumn{1}{l}{$w_{\text{c}}$} & \multicolumn{1}{l|}{$w_{\text{f}}$} & \multicolumn{1}{l}{$E_\text{d}$} & \multicolumn{1}{l}{$E_\text{c}$} & \multicolumn{1}{l|}{$E_\text{f}$} & \multicolumn{1}{l|}{iterations} & \multicolumn{1}{l}{$T_{\text{total}}$} & \multicolumn{1}{l}{$T_{\text{solver}}$} & \multicolumn{1}{l|}{$T_{\text{iter}}$} \\ \hline
    91    & 200   & 1073  & 100   & 0.01  & 0.1   & 2.54 & 1.8 & 0.74 & 60    & 121.76 & 0.13 & 2.03 \\ \hline
    \end{tabular}%
  \label{tab:leather}%
\end{table}%


\end{example}

\begin{example}\label{ex: simple developables}

In this example, we will focus on optimizing two relatively simple non-developable surfaces for planarity of their respective Gauss images. We start with two bicubic B\'ezier surfaces $\mathbf{S}^{a}_0$ and $\mathbf{S}^{b}_0$, where $\mathbf{S}^{a}_0$ is of mainly negative Gaussian curvature and $\mathbf{S}^{b}_0$ of positive Gaussian curvature.



\begin{figure}[H]
	\begin{tikzpicture}
	\node[anchor=south west,inner sep=0] (image) at (0,0) {\includegraphics[width=\linewidth]{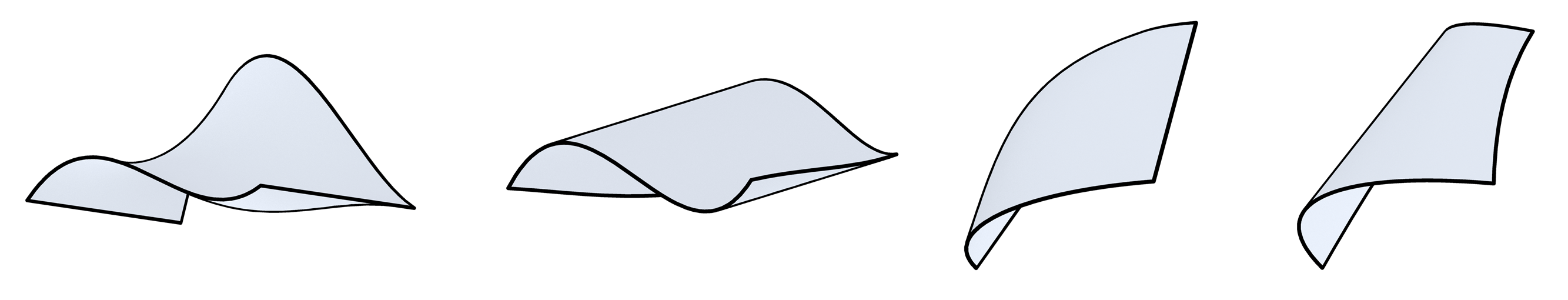}};
	\begin{scope}[x={(image.south east)},y={(image.north west)}]
	
	\draw[lightgray, thick] (0.595,0.05) -- (0.595,0.9);
	
	\node[above] at (0.06,0.7) {$\mathbf{S}^{a}_0$};
	\node[above] at (0.36,0.7) {$\mathbf{S}^{a}$};
	
	\node[above] at (0.63,0.75) {$\mathbf{S}^{b}_0$};
	\node[above] at (0.85,0.75) {$\mathbf{S}^{b}$};
	
	\node[above right] at (0.025,0) {\small\textbf{(a)}};
	\node[above left] at (0.975,0) {\small\textbf{(b)}};
	
	\draw[->, >=latex, line width=3pt, color=gray] (0.29,0.55) -- (0.31,0.55);
	\draw[->, >=latex, line width=3pt, color=gray] (0.795,0.55) -- (0.815,0.55);
	
	\end{scope}
	\end{tikzpicture}%
	\par\noindent\hrulefill
	\caption{The initial surfaces $\mathbf{S}^{a}_0$, $\mathbf{S}^{b}_0$ and the optimized surfaces $\mathbf{S}^{a}$, $\mathbf{S}^{b}$ are shown from an appropriate angle to better showcase the emergence of rulings in the direction of least absolute principal curvature on each of the surfaces.}
	\label{fig: simple developables}
\end{figure}
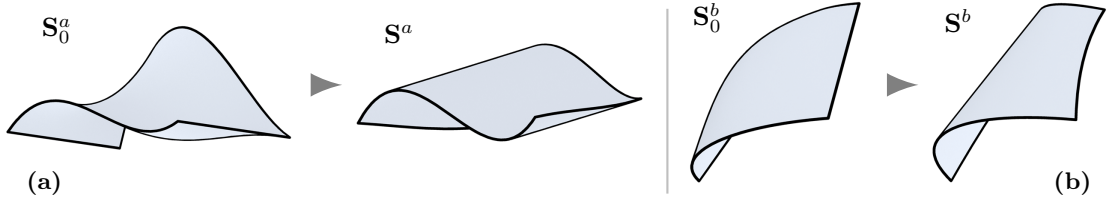


We follow optimization problem \ref{op: surface paneling optimization problem}, defined over a single panel, and utilize only the closeness and developability terms. Given that the surfaces have approximately planar Gauss images after the optimization, we also execute the following procedure at a point set $U$ on the surface to extrapolate the approximate rulings that are derived from their planar Gauss images, defined by the target plane $H$. We do this to present a visual comparison between these induced rulings and the computed rulings on the optimized surface. 

\begin{procedure}[H]
\caption*{\textbf{Procedure} Induced rulings}
\label{proc: principal directions}
\begin{algorithmic}
\ForAll{$\mathbf{p} \in U$}
	\State $\mathbf{n} \gets \sigma(\mathbf{p})$
	\State \textit{$\mathbf{q} \gets$ closest point of $\mathbf{n}$ to target circle $H\cap S^2$}
	\State \textit{$\mathbf{r}^t_\mathbf{q} \gets$ vector tangent to target circle at $\mathbf{q}$}
	\State \textit{$\mathbf{r}^o_\mathbf{q} \gets$ vector tangent to $S^2$ at $\mathbf{q}$ and orthogonal to $\mathbf{r}^t_q$} \Comment \textit{induced ruling direction}
	\State \textit{translate vectors $\mathbf{r}^t_q$, $\mathbf{r}^o_q$ to $\mathbf{p}$}
\EndFor
\end{algorithmic}
\end{procedure}

The vector $\mathbf{r}^o_q$ approximates the direction of the line generator of the surface at point $\mathbf{q}$. Moreover, for non-inflection rulings and non-planar regions on the optimized surfaces, vectors $\mathbf{r}^t_\mathbf{q}$, $\mathbf{r}^o_\mathbf{q}$ correspond to the principal directions of the surface at point $\mathbf{q}$.

Figure \ref{fig: simple developables} shows the surfaces before and after the optimization, while Figure \ref{fig: simple developables rulings} shows the resulting vectors from the \emph{Induced rulings} procedure.\\


\begin{figure}[t]
  \begin{tikzpicture}
  \node[anchor=south west,inner sep=0] (image) at (0,0) {\includegraphics[width=0.98\linewidth]{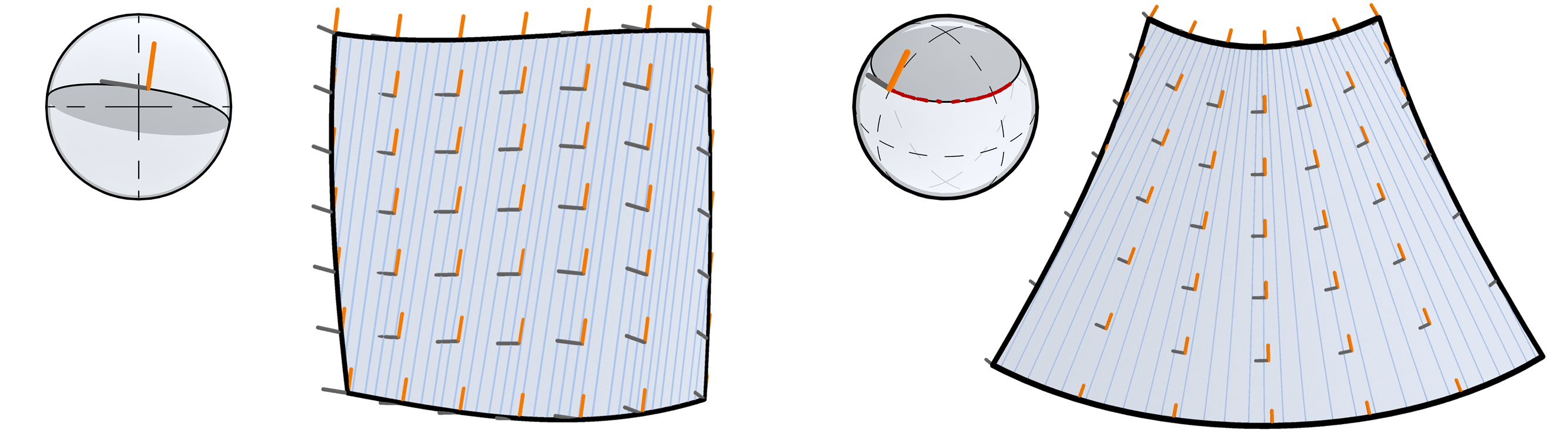}};
    \begin{scope}[x={(image.south east)},y={(image.north west)}]
      
      \draw[lightgray, thick] (0.505,0.05) -- (0.505,0.9);
      
      \draw[dashed,<->,thick] (0.105,0.85) -- (0.24,0.81);
      
      \draw[gray, line width=3pt, cap=round] (0.12,0.22) -- (0.07,0.22);
      \draw[orange, line width=3pt, cap=round] (0.12,0.22) -- (0.12,0.42);
      \node[right] at (0.12,0.42) {$\mathbf{r}^o$};
      \node[above] at (0.07,0.22) {$\mathbf{r}^t$};
      
      \node[left] at (0.2,0.19) {$\mathbf{S}^{a}$};
      
      \node at (0.6,0.3) {$\mathbf{S}^{b}$};
      
      \draw[dashed,-,thick] (0.045,0.74) -- (0.03,0.5);
      \node[below] at (0.029,0.5) {$H$};

      \node[above right] at (0,0) {\small\textbf{(a)}};
      \node[above] at (0.55,0) {\small\textbf{(b)}};
      

    \end{scope}
  \end{tikzpicture}%
  \par\noindent\hrulefill
\caption{A top-down perspective of the optimized surfaces $\mathbf{S}^{a}$, $\mathbf{S}^{b}$ is shown with the rulings superimposed on the surfaces (darker blue lines) as well as the resulting vectors from the predefined \emph{Induced rulings} procedure. We draw attention to the comparison between the orthogonal vectors $\mathbf{r}^o$ (orange) and the direction of the rulings (vanishing principal curvature direction). Furthermore, vectors $\mathbf{r}^t$ correspond to the directions of nonzero principal direction.}
\label{fig: simple developables rulings}
\end{figure}
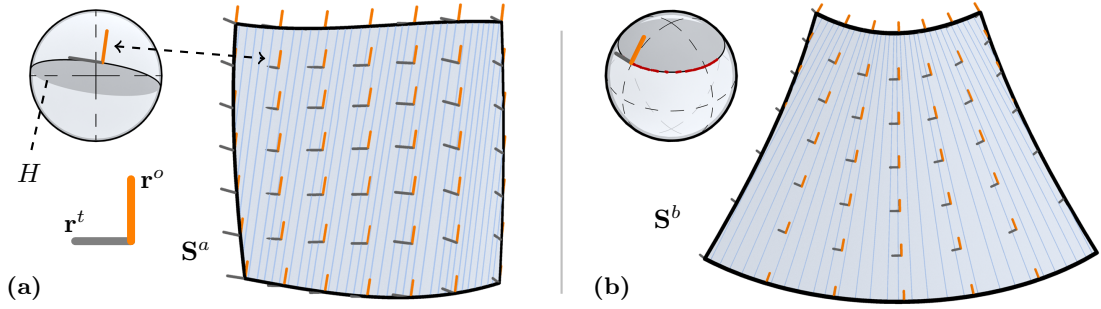


\setlength\tabcolsep{3pt}
\begin{table}[H]
  \centering
  \footnotesize
  \caption{The statistics for the Gauss image planarity optimization of panel surfaces $\mathbf{S}^{a}_0$ and $\mathbf{S}^{b}_0$. The weights were chosen to favor the developability property. Also provided, the total time, time used by the Newton solver, and the time per iteration (in seconds), measured on an Intel\textsuperscript{\tiny\textregistered} Core\textsuperscript{\tiny\texttrademark} i7-6700HQ processor.}
    \begin{tabularx}{\textwidth}{|Y|YYYY|YY|YY|c|YYY|}
    \hline
    \multicolumn{1}{|Y|}{Fig.} & \multicolumn{4}{c|}{number of...} & \multicolumn{2}{c|}{weights} & \multicolumn{2}{c|}{final energies} & \multicolumn{1}{c|}{number of} & \multicolumn{3}{c|}{time [sec]}\\
     \multicolumn{1}{|Y|}{No.} & \multicolumn{1}{c}{ctrl.pts} & \multicolumn{1}{c}{panels} & \multicolumn{1}{c}{variables} & \multicolumn{1}{c|}{eval.pts} & \multicolumn{1}{c}{$w_{\text{d}}$} & \multicolumn{1}{c|}{$w_{\text{c}}$} & \multicolumn{1}{c}{$E_{\text{d}}$} & \multicolumn{1}{c|}{$E_\text{c}$} & \multicolumn{1}{c|}{iterations} & \multicolumn{1}{c}{$T_{\text{total}}$} & \multicolumn{1}{c}{$T_{\text{solver}}$} & \multicolumn{1}{c|}{$T_{\text{iter}}$} \\ \hline
    \ref{fig: simple developables}a &  16 & 1  &  52 & 169 & $100$ & 1 &  0.65 &  81.28 & 10 &  1.9 & 0.1 & 0.19 \\
    \ref{fig: simple developables}b &  16 & 1  &  52 & 169 & $100$ & 1 &  1.45 &  99.21 & 10 &  2.05 & 0.02 & 0.2  \\\hline
    \end{tabularx}%
  \label{tab: simple developables}%
\end{table}%
\setlength\tabcolsep{6pt}


\end{example}

\begin{example}\label{ex: torus}

We provide here an introductory example of paneling a simple double curved surface with a variable number of rotational cylindrical panels.

We consider a surface $\mathbf{S}_{\text{ref}}$ which is a subset of the positive-Gaussian-curvature part of a torus. The active surface $\mathbf{S}$ of the optimization consists of a $N\times 1$ grid of bicubic panels. The initial configuration of $\mathbf{S}$ is given by fitting surface $\mathbf{S}$ to $\mathbf{S}_{\text{ref}}$.

We optimize for the panels of $\mathbf{S}$ to be rotational cylinders in the following manner. First of all, we use Lemma \ref{le: panel types} and assign to each panel an energy term of the form (\ref{eq: panel developability energy term}) with $d_r=0$ since we are interested in only cylindrical panels. We then solve optimization problem \ref{op: surface paneling optimization problem} with equal weights assigned to $E_\text{d}$ and $E_\text{r}$, and relatively smaller weights assigned to $E_\text{c}$ and $E_\text{f}$.

Figure \ref{fig: torus part} shows the resulting panelization for different values of $N$. We wish to direct the reader's focus to the curved boundary lines that follow the reference design; a characteristic not present and inherently not possible without trimming in previous approaches that utilized strips linear in one direction.\\


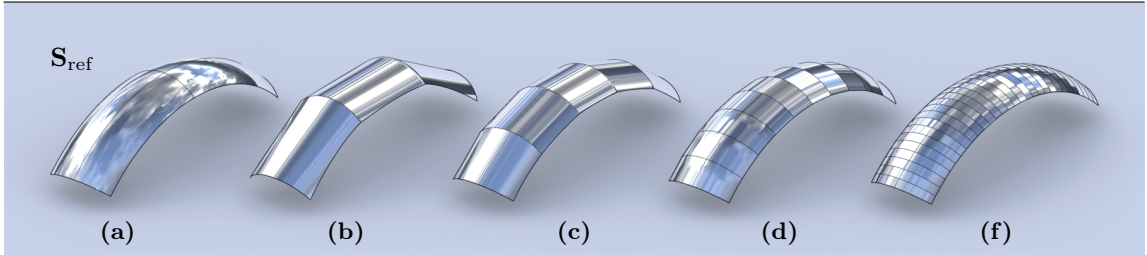
\begin{figure}[htbp]
\centering
\begin{tikzpicture}
  \node[anchor=south west,inner sep=0] (image) at (0,0) {\includegraphics[width=\linewidth]{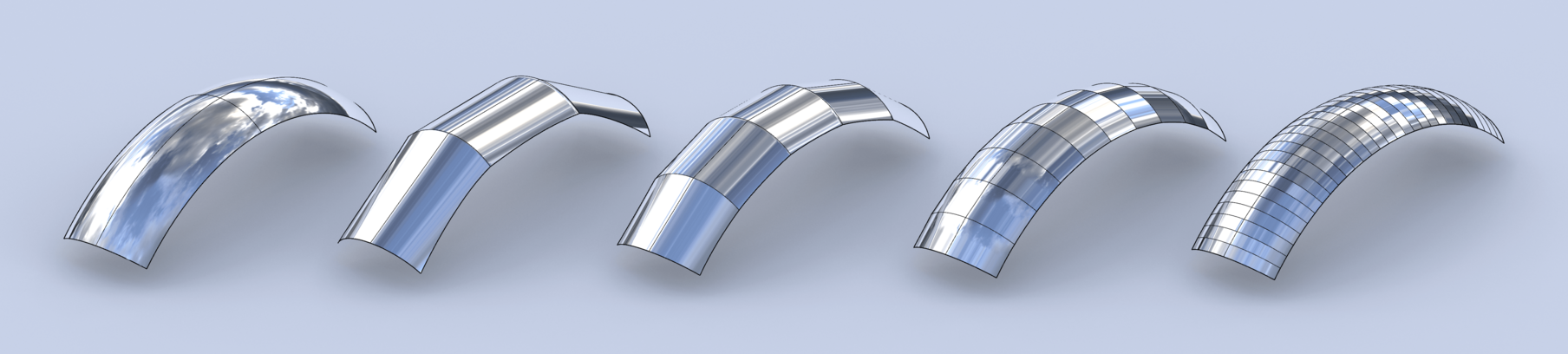}};
  \begin{scope}[x={(image.south east)},y={(image.north west)}]

  \node[above] at (0.06,0.7) {$\mathbf{S}_{\text{ref}}$};
  
  \draw[-] (0,1) -- (1,1);
  \draw[-] (0,0) -- (1,0);
  
  \node at (0.1,0.1) {\small\textbf{(a)}};
  \node at (0.3,0.1) {\small\textbf{(b)}};
  \node at (0.5,0.1) {\small\textbf{(c)}};
  \node at (0.68,0.1) {\small\textbf{(d)}};
  \node at (0.87,0.1) {\small\textbf{(f)}};
  
  \end{scope}
\end{tikzpicture}
\caption{Paneling part of a torus with a different number of cylindrical panels. Both the cutting planes $U_r$ per panel $\mathbf{S}^{(r)}$ and the inner boundary curves follow the direction of the smaller radius circles that define the torus.}
\label{fig: torus part}
\end{figure}


\setlength\tabcolsep{1.5pt}
\begin{table}[H]
  \centering
  \footnotesize
  \caption{The statistics for the paneling of the torus subsurface $\mathbf{S}_\text{ref}$ for different numbers of panels. Each panel was sampled uniformly at $4\times 4$ points for the developability term and at $10\times 10$ points for the closeness term. The weights were chosen to favor the developability property.  Also provided, the total time, time used by the Newton solver, and the time per iteration (in seconds), measured on an Intel\textsuperscript{\tiny\textregistered} Core\textsuperscript{\tiny\texttrademark} i7-6700HQ processor.}
    \begin{tabularx}{\textwidth}{|Y|YYY|YYYY|YYY|c|YYY|}
    \hline
    \multicolumn{1}{|Y|}{Fig.} & \multicolumn{3}{c}{number of...} & \multicolumn{4}{|c|}{weights} & \multicolumn{3}{c|}{final energies} & \multicolumn{1}{c|}{number of} & \multicolumn{3}{c|}{time [sec]}\\
     \multicolumn{1}{|Y|}{No.} & \multicolumn{1}{c}{ctrl.pts} & \multicolumn{1}{c}{panels} & \multicolumn{1}{c}{variables} & \multicolumn{1}{|c}{$w_{\text{d}}$} & \multicolumn{1}{c}{$w_{\text{r}}$} & \multicolumn{1}{c}{$w_{\text{c}}$} & \multicolumn{1}{c|}{$w_{\text{f}}$} & \multicolumn{1}{c}{$E_{\text{d}+\text{r}}^\dagger$} & \multicolumn{1}{c}{$E_\text{c}$} & \multicolumn{1}{c|}{$E_\text{f}$} & \multicolumn{1}{c|}{iterations} & \multicolumn{1}{c}{$T_{\text{total}}$} & \multicolumn{1}{c}{$T_{\text{solver}}$} & \multicolumn{1}{c|}{$T_{\text{iter}}$} \\ \hline
    \ref{fig: torus part}b &  40 & 3  &  132 & $10^2$ &  1 & 1 & 0.1 & 0.043  & 9.97 &  5.05 & 5 &  1.12 & 0.02 & 0.22 \\
    \ref{fig: torus part}c &  64 & 5  &  212 & $10^2$ &  1 & 1 & 0.1 & 0.004  & 2.09 &  5.01 & 5 &  2.05 & 0.03 & 0.4  \\
    \ref{fig: torus part}d & 124 & 10 &  412 & $10^3$ & 10 & 1 & 0.1 & 0.003  & 0.26 &  7.11 & 5 &  2.99 & 0.05 & 0.6  \\
    \ref{fig: torus part}e & 364 & 30 & 1212 & $10^3$ & 10 & 1 & 0.1 & 0.0002 & 0.05 & 18.68 & 5 &  8.11 & 0.17 & 1.62 \\\hline
    \multicolumn{15}{l}{\quad $^\dagger E_{\text{d}+\text{r}} = E_\text{d} + E_\text{r}$} \\
    \end{tabularx}%
  \label{tab:torus}%
\end{table}%
\setlength\tabcolsep{6pt}


\end{example}

\begin{example}

We extend the previous example of optimizing a simple row of panels to be of cylindrical type to the task of optimizing a grid of panels to be of any developable type we have previously addressed for panels.

Motivated by the possible architectural applications of the algorithm presented in this paper, we use as a reference surfac an architectural surface recently realized as the roof of the Department of Islamic Art at Musée du Louvre in Paris, France, shown in Figure \ref{fig: flying carpet}. The underlying surface is a highly non-developable surface with a strong variation in the sign of
Gaussian curvature. In this example, we set forth to compute an alternative realization of the same surface by using rotational conical and rotational cylindrical panels.\\


\begin{figure}[H]
\centering
\includegraphics[width=\linewidth]{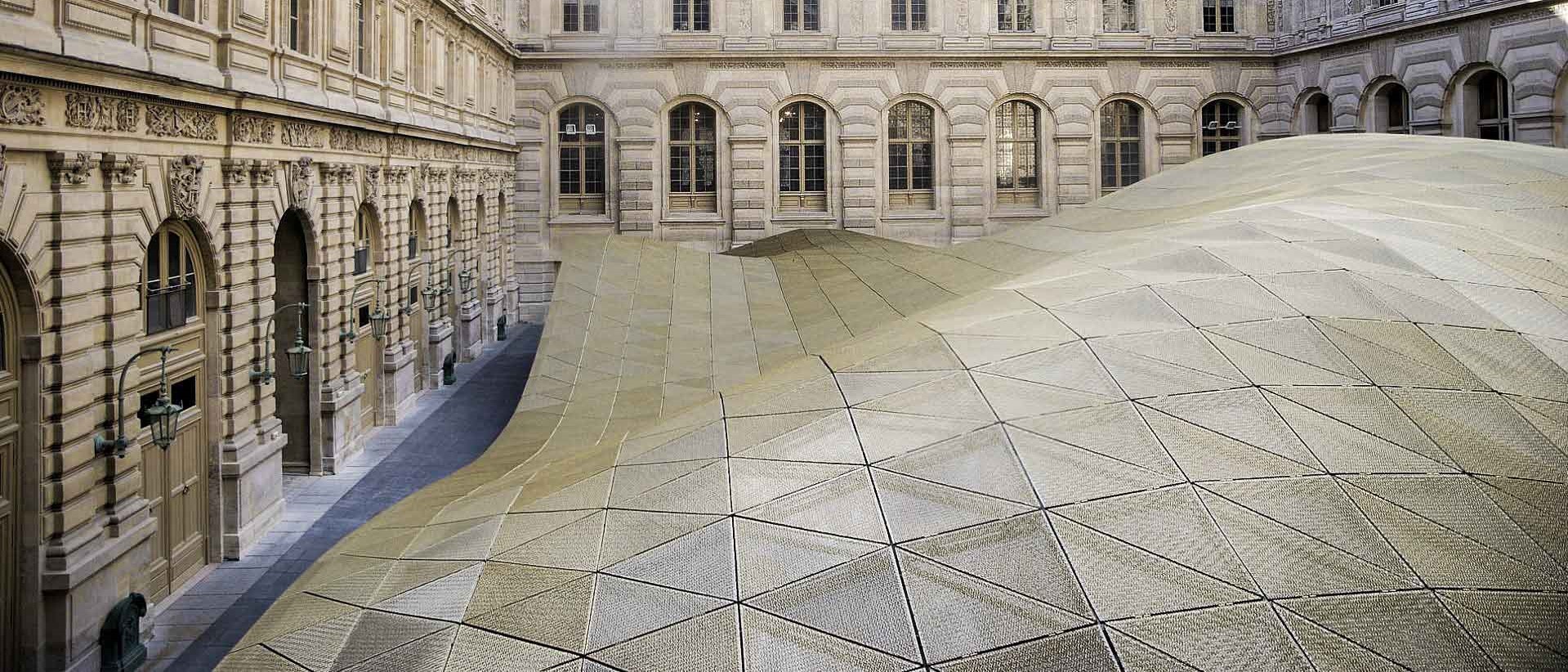}
\caption{Detail from the \emph{Cour Visconti} roof of the Department of Islamic Art at Musée du Louvre in Paris, France.}
\label{fig: flying carpet}
\end{figure}


The user input in this case is the freeform reference surface $\mathbf{S}_{\text{ref}}$, the desired number of panels in each direction of the grid that will constitute the panelization of the surface and the preferred type of panels, which includes surfaces of constant slope or the more specialized and more widely-used rotational surfaces of constant slope, i.e. rotational conical and rotational cylindrical. The user by adjusting the weights of the different energy terms involved in the corresponding optimization problem \ref{op: surface paneling optimization problem}, has influence over the various desirable aspects of the resulting panelization. In this particular example, we wish to use any of the types introduced before, namely rotational conical, rotational cylindrical and planar panels.

We present in Figure \ref{fig: flying carpet panelization} the resulting panelization of the reference surface for different panel grid resolutions. We set weight $w_{\text{c}}$, corresponding to the closeness of $\mathbf{S}$ to $\mathbf{S}_{\text{ref}}$, relatively high to reinforce the resulting surface to not deviate significantly from the reference surface and closely follow the chosen design. The smoothness of the boundary curves is controlled by the fairness energy term weight $w_{\text{f}}$, which is assigned a small value to ensure more visually pleasing results.\\


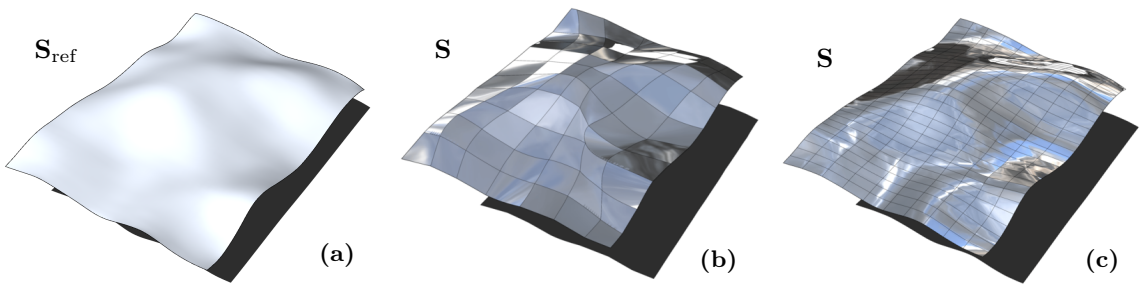
\begin{figure}[H]
\centering
\begin{subfigure}[b]{0.33\linewidth}
  \caption{}%
  \vspace{-16pt}
  \begin{tikzpicture}
  \node[anchor=south west,inner sep=0] (image) at (0,0) {\includegraphics[width=\textwidth]{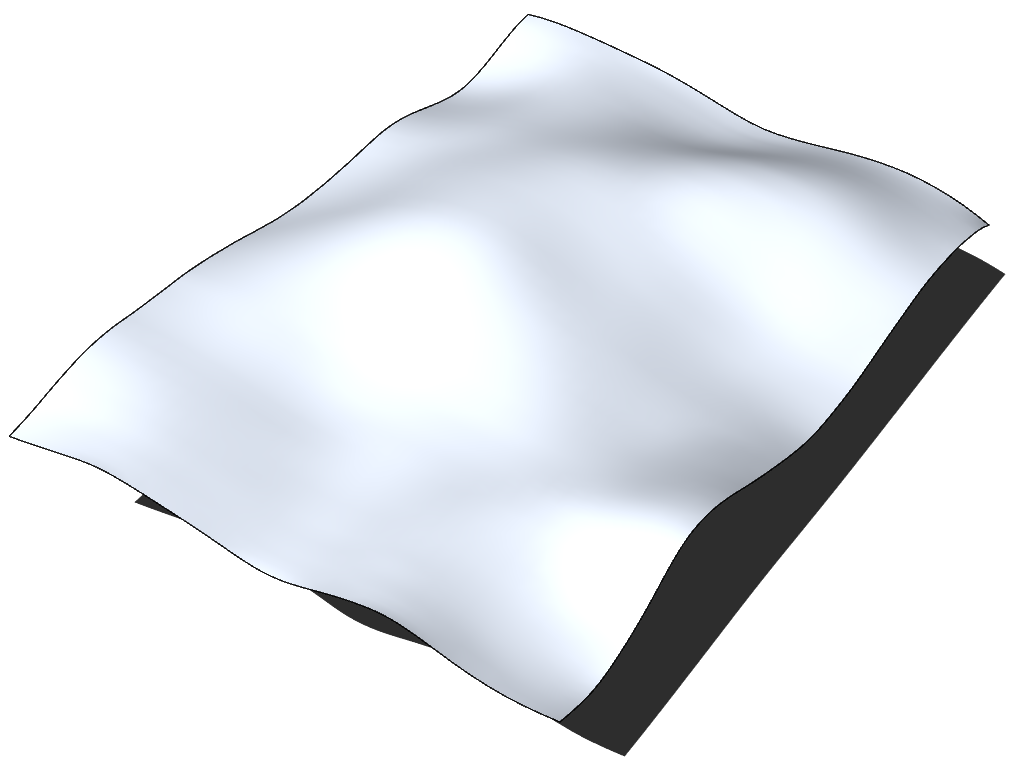}};
  \begin{scope}[x={(image.south east)},y={(image.north west)}]
  \node[below right] at (0.06,0.92) {$\mathbf{S}_{\text{ref}}$};
  \node[above left] at (0.96,0.04) {\small\textbf{(a)}};
  \end{scope}
  \end{tikzpicture}%
  \label{subfig: flying carpet}
\end{subfigure}%
\hfill%
\begin{subfigure}[b]{0.32\linewidth}
  \caption{}%
  \vspace{-16pt}
  \begin{tikzpicture}
  \node[anchor=south west,inner sep=0] (image) at (0,0) {\includegraphics[width=\textwidth]{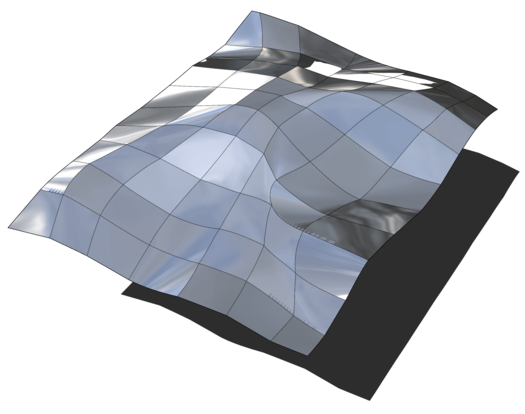}};
  \begin{scope}[x={(image.south east)},y={(image.north west)}]
  \node[below right] at (0.08,0.9) {$\mathbf{S}$};
  \node[above left] at (0.96,0.02) {\small\textbf{(b)}};
  \end{scope}
  \end{tikzpicture}%
  \label{subfig: flying carpet coarse panelization}
\end{subfigure}%
\hfill%
\begin{subfigure}[b]{0.32\linewidth}
  \caption{}%
  \vspace{-16pt}
  \begin{tikzpicture}
  \node[anchor=south west,inner sep=0] (image) at (0,0) {\includegraphics[width=\textwidth]{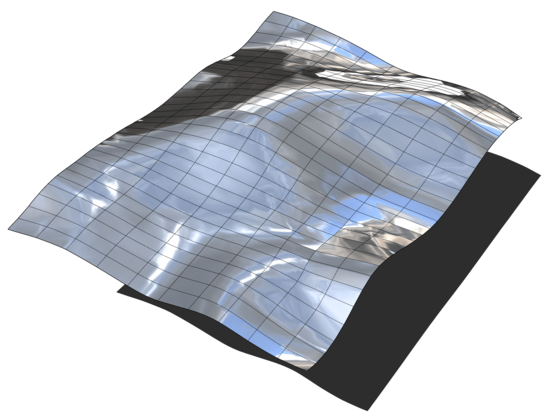}};
  \begin{scope}[x={(image.south east)},y={(image.north west)}]
  \node[below right] at (0.08,0.9) {$\mathbf{S}$};
  \node[above left] at (0.96,0.02) {\small\textbf{(c)}};
  \end{scope}
  \end{tikzpicture}%
  \label{subfig: flying carpet dense panelization}
\end{subfigure}%
\vspace{-5pt}
\par\noindent\hrulefill
\caption{\textbf{(a)} The freeform reference surface to be panelized. \textbf{(b)} A coarse panelization consisting of 70 panels. \textbf{(c)} A denser panelization consisting of 300 panels. Runtime for both the coarse and the finer paneling was several minutes.}
\label{fig: flying carpet panelization}
\end{figure}


The coarse panelization of Figure \ref{subfig: flying carpet coarse panelization} serves as a nice example of the dynamic panel layout adaptation which aims to approximate the given reference surface while satisfying the developability, rotationality and closeness constraints. On the contrary, by increasing the number of the panels utilized, we achieve the dense panelization of Figure \ref{subfig: flying carpet dense panelization}. As expected the increased number of panels produces an improved result, compared to the coarse equivalent. It not only better approximates the reference surface but also satisfies to a higher degree the additional secondary constraints, yielding a panelization of the reference surface that allows for a more structured arrangement of the panels. 

Nevertheless, both results are welcome since each one of them serves as a valid panelization with specialized developables of the same architectural surface. Each one of the two panelizations of this example shown in Figures \ref{subfig: flying carpet coarse panelization}, \ref{subfig: flying carpet dense panelization} manages to be architecturally aesthetically pleasing in its own style, while being realizable only by rotational cylindrical and rotational conical panels; highlighting the freedom of design expression that this method provides.
\end{example}

\smallskip
\noindent\textbf{Short discussion.} Appropriate choice of weights leads to high-precision satisfaction of the hard nonlinear constraints. The fairness and closeness terms act as regularizers to the optimization problem, which is formulated through simple polynomial energies. The combination of the soft constraints and fixed points, avoids degenerate results. The complexity of the approach is derived by the degree of the surface to be optimized, the reference surface (number of points of mesh representation) and number of evaluation points. In most applications, our experiments show that the computation time is limited to several seconds to get satisfactory results.

The presented local shaping approach achieves to minimize the predefined energies at every step, and guides iteratively the surface to an expected result. Any unwanted results were limited to surfaces that could not satisfy adequately both the closeness term and the developability term, meaning the result had to deviate considerably from the reference to satisfy the developability constraint.

\smallskip
\noindent\textbf{Limitations.} Among the limitations of our research, we first point to the lack of a material-dependent measure for the deviation from developability. The thickness of the Gauss image alone is not sufficient for judging whether a panel, fabricated from hardly stretchable material, can be easily bent into the computed shape. Moreover, our current implementation for paneling is limited to a grid type arrangement of panels and could benefit from additional improvements to the optimizer. 

\smallskip
\noindent\textbf{Conclusion.} 
We have introduced a methodology for increasing the developability of surfaces through an optimization algorithm which aims
at a thin Gauss image. Our implementation uses B-spline surfaces, but an analogous approach could be formulated for other 
surface representations as well. Moreover, we have presented a novel paneling algorithm which---in contract to prior
work \cite{pottmann-2010-pan}---optimizes both for the panels and the curve network of panel boundaries, under the constraint that
panels are developable with a planar Gauss image and/or rotational. 

\smallskip
\noindent\textbf{Future work.} A promising and important direction for future work is to
incorporate a specific material behavior. For example, it would be nice to come up with an efficient algorithm
that automatically enforces the design of only those surfaces which can easily be produced from a given material.
In particular, materials which bend much more easily than they stretch are of high interest. This leads into
the geometrically largely unexplored area of nearly developable surfaces. The paneling algorithm would greatly 
benefit from an extension to more general panel arrangements, maybe incorporating user interaction supported by
automatic suggestions of the system.


\section{Acknowledgements}

We would like to thank Heinz Schmiedhofer for providing the scan and picture of the leather surface example in Figure \ref{fig:leather}. This project has received funding from the European Union’s Horizon 2020 research and innovation programme under the Marie Skłodowska-Curie grant agreement No 675789.

\bibliographystyle{elsarticle-num}
\bibliography{bib}

\end{document}